\documentclass{ptephy}

\usepackage{amsmath,amssymb}
\usepackage{bm}
\usepackage{graphicx}
\usepackage{verbatim}
\usepackage{wrapfig}
\usepackage{ascmac}
\usepackage{makeidx}
\usepackage{mymacro}
\usepackage{color}
\usepackage{comment}

\makeatletter

\begin{document}

\title{Non-adiabatic effect in quantum pumping for a spin-boson system}

\author{\name{Kota L. Watanabe}{1\ast} and  \name{Hisao Hayakawa}{2\dagger}}
\address{\affil{1,2}{Yukawa Institute for Theoretical Physics, Kyoto University Kitashirakawa Oiwakecho, Sakyo-ku, Kyoto 606-8502 Japan}
\email{knabe@yukawa.kyoto-u.ac.jp, $^\dagger$E-mail: hisao@yukawa.kyoto-u.ac.jp}}

\begin{abstract}

We clarify the role of non-adiabatic effects in quantum pumping for a spin-boson system.
When we sinusoidally control the temperatures of two reservoirs with $\pi/2$ phase difference, we find that the pumping current strongly depends on the initial condition, and thus, the current deviates from that predicted by the adiabatic treatment. 
We also analytically obtain the contribution of non-adiabatic effects in the pumping current proportional to $\Omega^3$ where $\Omega$ is the angular frequency of the temperature control.
The validity of the analytic expression is verified by our numerical calculation.
Moreover, we extend the steady heat fluctuation theorem to the case for slowly modulated temperatures and large transferred energies.
\end{abstract}

\maketitle

\section{Introduction}
A pump converts an external bias into work.
We need the average bias to get the work from a macroscopic mechanical pump, 
but it is known that  the average bias to get a pumping current is not necessary in mesoscopic systems.
When a mesoscopic system, thus, is slowly and periodically modulated by several control parameters such as chemical potentials, gate voltages, and tunneling barriers, there exists a net average current without dc bias.
This phenomenon is known as adiabatic pumping, and has been observed in various processes such as quantized charge transport\cite{Thouless,Niu,Avron,Kouwenhoven,Pothier,Fuhrer,Kaestner,Chorley,Andreev,Makhlin,Aleiner}, spin pumping\cite{Mucciolo,Governale,Cota,Splettstoesser1,Riwar,Deus,Watson}, and qubit manipulation\cite{Brandes}.
The first proposal of adiabatic pumping was given by Thouless\cite{Thouless} for a closed quantum system.
The idea of quantum pumping for closed systems has been extended to open systems\cite{Andreev,Aleiner,Buttiker1,Buttiker2,Brouwer1,Zhou,Cremers,Moskalets,Brouwer2,Stefanucci,Breuer}. 
Such adiabatic pumping processes have been experimentally realized in mesoscopic transport processes\cite{Kouwenhoven,Pothier,Fuhrer,Kaestner,Chorley,Watson,Tsukagoshi,Buitelaar,Switkes,Giazotto}. 
It is recognized that the mechanism of adiabatic pumping originates from the geometrical effect of the Berry phase in quantum mechanics\cite{Berry},
where a circular operation in a parameter space creates a non-zero geometrical quantity associated with the pumping current.

Similar phenomena have been studied in stochastic systems described by classical master equations\cite{Parrondo,Usami,Astumian1,Sinitsyn1,Sinitsyn2,Astumain2,Rahav,Ohkubo,Ren,Sagawa,Chernyak1,Chernyak2} and quantum master equations\cite{Cota,Splettstoesser1,Riwar,Brandes,Renzoni,Splettstoeser2,Reckemann,Hiltscher,Yuge,Yoshii}.
As indicated in the analysis of classical master equations\cite{Breuer,Parrondo,Usami,Astumian1,Sinitsyn1,Sinitsyn2,Rahav,Ohkubo,Ren,Sagawa},  adiabatic pumping is also characterized by a Berry-phase-like quantity, the so-called Berry-Sinitsyn-Nemenman (BSN) phase\cite{Sinitsyn1,Sinitsyn2,Ohkubo,Ren,Sagawa}.
The BSN phase has been extended to the quantum master equation case\cite{Yuge}.
It is remarkable that the BSN phase is directly related to the path-dependent entropy under strong nonequilibrium conditions\cite{Sagawa,Yuge13}, which is an interesting extension of the equilibrium thermodynamics to a nonequilibrium thermodynamics.

Most of the previous studies, however, assume that the pumping process is only modulated adiabatically, 
where the validity of the approximation is ensured if the modulation speed is zero.
This situation is practically useless, because the pumping current under adiabatic modulation is zero in the strict sense.
It is, thus, important to (i) clarify the limitation of the adiabatic approximation and (ii) analyze the pumping process without the introduction of the adiabatic approximation to get a finite pumping current under a finite speed modulation.

Although there exist some papers discussing non-adiabatic pumping effects based on a stochastic equation with weak noise\cite{Strass}, the master equation\cite{Uchiyama},  the Floquet scattering theory\cite{Moskalets2008}, and the Green function\cite{Wang,Arrachea}, it is unclear how non-adiabatic effects affect the pumping current.
Indeed, it is known that a non-adiabatic process can cause a phase transition through the analysis of a simple quantum mechanical model\cite{Sinitsyn13}. 

We may ask another non-trivial question associated with the non-adiabatic pumping process besides the pumping current. 
Although there exists the heat fluctuation theorem\cite{Esposito,Jarzynski04,Saito07,Talkner,Noh,Gaspard07}, for adiabatic dynamics of open Markovian processes, at least, the heat fluctuation theorem seems to be violated under some situations such as the dynamics under modulated external fields\cite{Ren}, non-Gaussian noise\cite{kanazawa13}, or dry friction\cite{sano}. 
We have to clarify the reason why the heat fluctuation theorem seems to be violated.

In this paper, we systematically study non-adiabatic pumping effects within the framework of the quantum master equation under the Markovian approximation.
For this purpose, similar to Ref.\cite{Uchiyama},  we analyze the simplest spin-boson model under the weak coupling condition between surrounding environments and the system.
We continuously control the temperatures in the environments with the modulation frequency $\Omega/2\pi $, and clarify the initial condition dependence of the pumping current and the essential non-adiabatic effects on the pumping current. 
We also extend the steady heat fluctuation theorem to cases of slowly modulated temperatures and high transferred energy limits. 

The organization of this paper is as follows.
In Sect.\ 2, we introduce the model of the spin-boson system and the methods of the generalized quantum master equation with the full counting statistics (FCS). 
Section 3 is the main part of this paper, and consists of three parts. 
In Sect.\ 3.1, we derive general expressions for the non-adiabatic pumping current. 
In Sect.\ 3.2, we apply our formulation to the spin-boson system introduced in Sec.\ 2, and present the results for the pumping current to clarify the non-adiabatic effects.
In Sect.\ 3.3, we discuss whether the heat fluctuation theorem is still valid. 
Finally, we discuss and summarize our results in Sect.\ 4. 
In Appendix A, we briefly summarize the properties of the cumulant-generating function and the first moment.
In Appendix B, we derive the master equation with parameter modulation in the context of FCS.
In Appendix C, we reproduce the adiabatic Markovian pumping current obtained in Ref.\cite{Ren} within our framework.
In Appendix D, we summarize the relationship between our formulation and that in Ref.\cite{Ren}.
In Appendix E, we derive the asymptotic expansion of the density matrices and the non-adiabatic pumping current. 
In Appendix F, we explain the detailed derivation of the extended heat fluctuation theorems showed in Sect.\ 3.3.

\section{Model and method}

In this section, we introduce our model and the method to be used in our analysis. 
We analyze a spin-boson system, and adopt  the generalized quantum master equation with the full counting statistics (FCS) as the basic equation for our analysis. 

The spin-boson system is a simple two-level system $\{ \ket 0, \ket 1\}$ coupled with two environments (denoted as $L$ and $R$) characterized by the inverse temperatures $\beta_\nu$ where $\nu=L$ or $R$.
We modulate the temperatures periodically with the angular frequency $\Omega$ under the condition that the environments are always in equilibrium.
The system Hamiltonian $H_{\rm S}$ and the environmental Hamilitonian $H_{\rm E}^\nu$ ($\nu=L$ or $R$) are, respectively, given by
\begin{equation}\label{H_s&H_E}
H_{\rm S}=\sum_{n=0,1}\epsilon_n |n \rangle \langle n| ,
\quad
H_{\rm E}^\nu=\sum_k \hbar \omega_{k,\nu}b_{k,\nu}^\dagger b_{k,\nu} ,
\end{equation}
where $b_{k,\nu}$ and $b_{k,\nu}^\dagger$ are, respectively, bosonic annihilation and creation operators at the wave number $k$ for the environment $\nu$, and
$\epsilon_n$ and $\omega_{k,\nu}$ are the energy for the level $n(=0,1)$ and the angular frequency characterizing the bosonic environment $\nu$, respectively. 
We introduce the characteristic frequency $\omega_0$ from the relation $\hbar \omega_0\equiv \epsilon_1-\epsilon_0 $. 
The interaction Hamiltonian $H_{\rm SE}^\nu$ is given by
\begin{equation}\label{H_{SE}}
H_{\rm SE}^\nu= \hbar (\ket{0}\bra{1}+\ket{1}\bra{0})\sum_k g_{k,\nu}(b_{k,\nu}+b_{k,\nu}^\dagger)
\end{equation} 
with the coupling strength $g_{k,\nu}$, which is characterized by the spectral density function $\Gamma_{\nu}(\omega) = 2\pi \sum_k g_{k,\nu}^2 \delta (\omega -\omega_{k,\nu})$. 
We assume that the environments are always characterized by the equilibrium operator $\rho^{\rm eq}_{\rm E}(\beta_{\nu})=e^{-\beta_{\nu}H_{\rm E}^{\nu}}/Z$.

To calculate the average energy transfer $\Delta q_t$ from a reservoir to the system during the time interval $t$, we use the FCS method. 
When the two-point projective measurement on a quantity $Q$ is performed at times $0$ and $t$, the corresponding outcomes are $q_0$ and $q_t$ respectively.
Thanks to the method of FCS, we can calculate the cumulant-generatig function $S(\chi,t) \equiv \ln{ \int P(\Delta q_t)e^{i\chi \Delta q_t}d\Delta q_t }$,
 where $P(\Delta q_t)$ is the probability distribution function of $\Delta q_t=q_t-q_0$ and $\chi$ is the counting field.
Once we know $S(\chi,t)$,  we can get the $n$th cumulant of $P(\Delta q_t)$ from the $n$th derivative of $S(\chi,t)$ at $\chi=0$.
Therefore the average energy transfer is given by $\expec{\Delta q_t}_c = \partial S(\chi,t)/\partial (i\chi) |_{\chi=0}$. 
The detailed method of the calculation of the cumulant-generating function $S(\chi,t)$ is explained  in Appendix A.
In this method, the cumulant-generating function is given by $S(\chi,t) =\tr\rho_{\rm tot}(\chi,t)$, where $\rho_{\rm tot}(\chi,t)$ is the generalized density matrix for the total system defined in Eq.\ (\ref{modrho}).

In the weak coupling limit $g_{k,\nu}\ll \omega_{k,\nu}, \epsilon_n/\hbar$, it is straightforward to obtain the quantum master equation for the reduced density matrix $\rho(\chi,t)\equiv {\rm Tr}_{\rm E} \rho_{\rm tot}(\chi,t)$ (see Appendix B). According to Appendix B, the correlation timescale $\tau_{\rm C}$ of environments is characterized by the symmetrized time correlation function, which is, for the operator of environments $B_{\nu}=\sum_k g_{k,\nu}b_{k,\nu}$ in our model, given by (see Ref.\cite{weiss})
\begin{eqnarray}
\Re\tr_{\rm E}[\{B_{\nu}(\tau)^{\dagger},B_{\nu}\}\rho^{\rm eq}_{\rm E}(\beta_{\nu})] &=& \int_0^{\infty}d\omega \Gamma_{\nu}(\omega)\frac{e^{\beta_{\nu}\hbar \omega_0}+1}{e^{\beta_{\nu}\hbar \omega_0}-1}\cos{(\omega \tau)} \nonumber \\
&=& \frac{g\omega_{c,\nu}^2}{2\pi}\frac{1-\tau^2\omega_{c,\nu}^2}{(1+\tau^2\omega_{c,\nu}^2)^2}+\frac{g}{\pi(\hbar \beta_{\nu})^2}\Re[\psi'(1+\frac{1}{\beta_{\nu}\hbar\omega_{c,\nu}}+i\frac{\tau}{\hbar\beta_{\nu}})], \nonumber \\ \label{tco}
\end{eqnarray}
where we have used the Bose distribution $\expec{b_{k,\nu}^{\dagger}b_{k,\nu}} = (e^{\beta_{\nu}\hbar \omega_{k,\nu}}-1)^{-1}$ and the Ohmic spectral density $\Gamma_\nu(\omega)=g\omega e^{-\omega/\omega_{c,\nu}}$ with the cutoff $\omega_{c,\nu}$. Here, $\Re A$ represents the real part of $A$ and $\psi(x)\equiv \Gamma'(x)/\Gamma(x)$ is the digamma function. For our setting of parameters in this paper, the characterized timescale $\tau_{\rm C}$ in Eq.\ (\ref{tco}) satisfies $\tau_{\rm C}\omega_0 \sim \beta_{\nu} \hbar \omega_0 \sim O(1)$. On the other hand, the relaxation timescale $\tau_{\rm R}$ of the system is estimated as $\tau_{\rm R}\omega_0 \sim 10^3$ and we consider $\tau_{\rm R} \lesssim \Omega^{-1}$. Therefore, if the condition $\tau_{\rm C} \ll \tau_{\rm R} \lesssim \Omega^{-1}$ is satisfied, we can derive the Markovian quantum master equation
\begin{eqnarray}\label{QME1}
\frac{d}{dt}\rho(\chi,t)=-\frac{i}{\hbar}[H_{\rm S},\rho(\chi,t)]  -\sum_{\nu=L,R}\frac{1}{\hbar^2}\int_0^{\infty} d\tau {\rm Tr}_{\rm E} [ H_{\rm SE}^\nu, [H_{\rm SE}^\nu(-\tau),\rho_{\rm E}({\bm \beta}(t)) \rho(\chi,t)]_\chi ]_\chi , 
\end{eqnarray}
where $[H,A]\equiv H_\chi A-A H_{-\chi}$ for an arbitrary operator $A$ and $H_\chi\equiv e^{i\chi Q/2}H e^{-i\chi Q/2}$, and $\bm{\beta}(t)$ is the vector representation of $\{\beta_\nu(t) \}$.
In the Markovian case, the spectral density $\Gamma_{\nu}(\omega)$ is reduced to the constant tunneling rate $\Gamma_{\nu}\equiv \Gamma_{\nu}(\omega_0)$. Because we consider identical environments, let us introduce $\Gamma \equiv \Gamma_L=\Gamma_R$, which characterizes the relaxation timescale $\tau_{\rm R} \sim \Gamma^{-1}$ of the system.

Let $\kket{\rho(\chi,t)}$ be the vector  $\kket{\rho(\chi,t)}\equiv {}^T(\langle 0| \rho(\chi,t)| 0\rangle, \langle 1| \rho(\chi,t)|1\rangle ) $ 
consisting of the diagonal element of $\rho(\chi,t)$ with the notation of the transverse  ${}^T\bm{A}$ of an arbitrary vector $\bm{A}$.
Note that the diagonal part of Eq.\ (\ref{QME1}) can be independent of the off-diagonal part in our model. 
Thus, the quantum master equation (\ref{QME1}) can be written as
\begin{eqnarray}\label{QME3}
\df{}{t}\kket{\rho(\chi,t)} = \mathcal{K}_M^{\chi}(\bm{\beta}(t)) \kket{\rho(\chi,t)},
\end{eqnarray}
where the evolution matrix $\mathcal{K}_M^\chi(\bm{\beta}(t))$ is given by
\begin{equation}\label{Markov-K}
\mathcal{K}_M^\chi(\bm{\beta}(t))
=-\int_0^\infty d\tau 
\left(
\begin{array}{cc}
 \zeta_1(\bm{\beta}(t),\tau) & \zeta_2^\chi(\bm{\beta}(t),\tau) \\
 \zeta_3^\chi(\bm{\beta}(t),\tau) & \zeta_4^\chi(\bm{\beta}(t),\tau)   \\
 \end{array}
\right) .
\end{equation}
Here, we have introduced
\begin{eqnarray}
\zeta_1(\bm{\beta},\tau) &=&  \sum_{\nu=L,R} \{ \Phi_{1,\nu}(\bm{\beta},\tau)e^{-i\omega_0\tau} + \Phi_{1,\nu}^{*}(\bm{\beta},\tau)e^{i\omega_0\tau}\}, \label{zeta1}\nonumber \\ \\
\zeta_2^{\chi}(\bm{\beta},\tau) &=&  -\sum_{\nu=L,R} \{ \Phi_{2,\nu}^{\chi}(\bm{\beta},\tau)e^{-i\omega_0\tau} + \Phi_{3,\nu}^{\chi}(\bm{\beta},\tau)e^{i\omega_0\tau}\}, \nonumber \\ \\
\zeta_3^{\chi}(\bm{\beta},\tau) &=&  -\sum_{\nu=L,R} \{ \Phi_{2,\nu}^{\chi}(\bm{\beta},\tau)e^{i\omega_0\tau} + \Phi_{3,\nu}^{\chi}(\bm{\beta},\tau)e^{-i\omega_0\tau}\}, \nonumber \\ \\
\zeta_4(\bm{\beta},\tau) &=&  \sum_{\nu=L,R} \{ \Phi_{1,\nu}(\bm{\beta},\tau)e^{i\omega_0\tau} + \Phi_{1,\nu}^{*}(\bm{\beta},\tau)e^{-i\omega_0\tau}\}, \label{zeta4}\nonumber \\
\end{eqnarray}
where 
\begin{eqnarray}
\Phi_{1,\nu}(\bm{\beta},\tau)&=&\sum_{k} g_{k,\nu}^2 \{ \expec{b_{k,\nu}^{\dagger}b_{k,\nu}}_{\bm{\beta}}e^{i\omega_{k,\nu}\tau} + \expec{b_{k,\nu}b_{k,\nu}^{\dagger}}_{\bm{\beta}}e^{-i\omega_{k,\nu}\tau}\}, \\
\Phi_{2,\nu}^{\chi}(\bm{\beta},\tau)&=
&\sum_{k} g_{k,\nu}^2
\{ \expec{b_{k,\nu}^{\dagger}b_{k,\nu}}_{\bm{\beta}}
e^{-i\omega_{k,\nu}\tau - i\hbar\omega_{k,\nu}\chi_{\nu}}
 + \expec{b_{k,\nu}b_{k,\nu}^{\dagger}}_{\bm{\beta}}e^{i\omega_{k,\nu}\tau + i\hbar\omega_{k,\nu}\chi_{\nu}}\}, \\
\Phi_{3,\nu}^{\chi}(\bm{\beta},\tau) &=&\sum_{k} g_{k,\nu}^2\{\expec{b_{k,\nu}^{\dagger}b_{k,\nu}}_{\bm{\beta}}e^{i\omega_{k,\nu}\tau - i\hbar\omega_{k,\nu}\chi_{\nu}} + \expec{b_{k,\nu}b_{k,\nu}^{\dagger}}_{\bm{\beta}}e^{-i\omega_{k,\nu}\tau + i\hbar\omega_{k,\nu}\chi_{\nu}}\}.
\end{eqnarray}
Here, we explicitly write the control parameters and inverse temperatures $\bm{\beta}$;
$\langle \cdot \rangle_{\bm{\beta}}$ represents the average over the bosonic field in the environment characterized by $\bm{\beta}$.
Namely, we have assumed that the environments are always in thermal equilibrium even if we modulate $\bm{\beta}(t)$. Thus, it is not appropriate to apply our formulation to too-fast modulations.
This means that we cannot use our theory for cases of abrupt temperature change.
We also assume that the time evolution of $\bm{\beta}(t)$ satisfies
\begin{eqnarray}\label{T(t)}
T_L(t)&=&  T_{0}+ T_A \cos(\Omega t+\pi/4) , \nonumber\\
T_R(t)&=&  T_{0}+ T_A \sin(\Omega t+\pi/4) ,
\end{eqnarray}
where $T_{0}$ and $T_A$ are, respectively, the average temperature and the amplitude of the modulation.

\section{ Main results : Non-adiabatic Markovian pumping}
\subsection{General expression}
It is straightforward to extend the adiabatic approximation used in Appendix C which is reduced to that used in Ref.\cite{Ren}. 
At first, let us decompose the average current into two parts:
\begin{eqnarray}\label{q_t}
\expec{\Delta q_t} &=& \expec{\Delta q_t}_{\rm na}^{\rm E}+\expec{\Delta q_t}^{\rm d},
\end{eqnarray}
which is a natural extension of Eq.\ (\ref{adiabatic_current}).
It should be noted that the contribution from the dynamical phase $\expec{\Delta q_t}^{\rm d}$ is invariant even in the non-adiabatic treatment, while
the adiabatic geometrical current in Eq.(\ref{adiabatic_geo_q}) is now replaced by $\expec{\Delta q_t}_{\rm na}^{\rm E}$ :
\begin{eqnarray}\label{nageo}
\expec{\Delta q_t}_{\rm na}^{\rm E} = - \int_{0}^td\tau \bbra{l_+'(\bm{\beta}(\tau))}\df{}{\tau}\kket{\rho(0,\tau)},
\end{eqnarray}
where $\bbra{l'_+}$ is the $\chi$-derivative at $\chi=0$ of $\bbra{l^{\chi}_+}$, which is the left eigenvector of $\mathcal K^{\chi}_M$ for the eigenvalue $\lambda^{\chi}_+$ with the maximum real part.
Namely, the right eigenvector $\kket{\lambda_+^0}$ for the steady state used for the adiabatic process is relpaced by the density matrix $\kket{\rho(0,t)}$.
This result can be interpreted as follows.
Because the dynamical phase depends only on the average bias for the symmetric cyclic modulation, it is reasonable that the pumping current through the dynamical phase is unchanged, even if we consider the non-adiabatic effects.
On the other hand, the adiabatic transfer $\expec{\Delta q}_{\rm a}^{\rm g}$ depends on modulation speed and the path on the parameter space. In the non-adiabatic case, hence, the excess energy transfer corresponding to the contribution from the geometrical phase in the adiabatic limit has to be replaced by $\expec{\Delta q_t}_{\rm na}^{\rm E}$.

Now, let us prove the expressions (\ref{q_t}) and (\ref{nageo}). The formal solution of Eq.\ (\ref{QME3}) is 
\begin{eqnarray}\label{formal}
\kket{\rho(\chi,t)} = T_{\rightarrow}\exp{\left( \int_0^td\tau\mathcal{K}_M^\chi(\bm{\beta}(\tau))\right)} \kket{\rho(\chi,0)},
\end{eqnarray}
where we have introduced the time-ordering product from left to right as $T_{\rightarrow} \exp[\int_0^td\tau \mathcal{ K}(\tau)]\equiv$
$\sum_{n=0}^{\infty} \int_{0}^t d s_1\int_0^{s_1} d s_2 \cdots \int_0^{s_{n-1}} d s_{n} \mathcal{K}(s_1) \mathcal{ K}(s_2)\cdots \mathcal{ K}(s_n)$.  
Thus, we obtain the expression of the energy transfer $\expec{\Delta q_t}$ as :
\begin{eqnarray}\label{pumtot}
\expec{\Delta q_t}= \int_0^td\tau\bbra{1} \mathcal{K}_M'(\bm{\beta}(\tau))\kket{\rho(0,\tau)},
\end{eqnarray}
where we have used $\expec{\Delta q_0}=0$ and the following deformation under the condition $\bbra{1}\mathcal{K}_M^0(\bm{\beta}(\tau))=0$ :
\begin{eqnarray}\label{deform}
& &\left. \pdf{}{(i\chi)}\bbra{1} T_{\rightarrow}\exp{\left( \int_0^td\tau\mathcal{K}_M^\chi(\bm{\beta}(\tau))\right)} \right|_{\chi=0} \nonumber \\
&=& \int_0^td\tau \bbra{1} \mathcal{K}_M'(\bm{\beta}(\tau))T_{\rightarrow}\exp{\left( \int_0^{\tau}d\tau'\mathcal{K}_M^0(\bm{\beta}(\tau'))\right)}. 
\end{eqnarray}
Equation (\ref{pumtot}) can be rewritten as 
\begin{eqnarray}\label{33}
\expec{\Delta q_t} &=& \int_0^td\tau \{ \bbra{l_+^{\chi}(\bm{\beta}(\tau))} \mathcal{K}_M^{\chi}(\bm{\beta}(\tau))\kket{\rho(\chi,\tau)} \}' \nonumber \\
& &- \int_0^td\tau \bbra{l_+'(\bm{\beta}(\tau))} \mathcal{K}_M^0(\bm{\beta}(\tau))\kket{\rho(0,\tau)}.
\end{eqnarray}
The first term on the right-hand side (RHS) of this equation is equal to  $\expec{\Delta q_t}^{\rm d}$ by using $\lambda_+^0=0$, 
and the second term on the RHS of (\ref{33}) is reduced to Eq.(\ref{nageo}) with the aid of Eq.(\ref{QME3}). 
Thus, we reach Eq.(\ref{q_t}).

\subsection{Application to the spin-boson System}

We now apply our formulation to the spin-boson system. Let us introduce $a^\chi_i(t) \equiv -\int_0^\infty d\tau \zeta_i^\chi(\bm{\beta}(t),\tau)\  (i=1,..., 4)$ where $\zeta_i^\chi(\bm \beta(t),t)$ is given in Eqs.\  (\ref{zeta1})--(\ref{zeta4}). 
We consider the case that the measured quantity is the Hamiltonian in the right environment $Q = H_{\rm E}^R$. Thus, the explicit form of each $a_i^{\chi}(t)$ in this case is given by:
\begin{eqnarray}
a_1(t) &=& -\Gamma_{L}n_{L}(t)-\Gamma_{R}n_{R}(t), \label{a1} \\
a_2^{\chi}(t) &=& \Gamma_{L}(1+ n_{L}(t))+\Gamma_{R}(1+ n_{R}(t))e^{i\chi \hbar \omega_0},\label{a2} \\
a_3^{\chi}(t) &=& \Gamma_{L}n_{L}( t)+\Gamma_{R}n_{R}(t)e^{-i\chi\hbar \omega_0},\label{a3} \\
a_4(t) &=&- \Gamma_{L}(1+ n_{L}(t))-\Gamma_{R}(1+ n_{R}(t)), \label{a4}
\end{eqnarray}
where $n_{\nu}(t)=(e^{\beta_\nu(t) \hbar \omega_0}-1)^{-1}$.
The eigenvalues and the eigenstates of $\mathcal{K}_M^\chi(\bm{\beta}(t))$ are explicitly written as
\begin{eqnarray}
\lambda^{\chi}_{\pm} (t) &=& \frac{a_1(t)+a_4(t)}{2} \pm \sqrt{\left(\frac{a_1(t)-a_4(t)}{2}\right)^2 + a^{\chi}_2(t) a^{\chi}_3(t) }, \label{lambda} \\
\label{39}
\kket{\lambda^{\chi}_{\pm}(t) } &=& \frac{1}{N_{\pm}^{\chi}(t)} \begin{pmatrix}
1 \\
\displaystyle \frac{\lambda^{\chi}_{\pm} (t)-a_1(t)}{a_2^{\chi}(t)}
\end{pmatrix} , \\
\bbra{l^{\chi}_{\pm}(t)} &=& \begin{pmatrix}
1 & \displaystyle \frac{\lambda^{\chi}_{\pm} (t)-a_1(t)}{a_3^{\chi}(t)}
\end{pmatrix}, \label{lefte}
\end{eqnarray}
where we have introduced
\begin{eqnarray}
N_{\pm}^{\chi}(t) = 1+ \frac{(\lambda^{\chi}_{\pm} (t)-a_1(t))^2}{a_2^{\chi}(t)a_3^{\chi}(t)} 
\end{eqnarray}
in Eq.\ (\ref{39}).
These expressions satisfy the orthonormal condition $\bbkk{l^{\chi}_i}{\lambda^{\chi}_j} = \delta_{ij}$．

When the counting field $\chi$ is absent, the above results reduce to
\begin{eqnarray}
&\lambda_{+}^{0}(t) = 0, \ \ \ \lambda_-^0(t) = \lambda(t),& \label{0lam}\\
&\kket{\lambda^{0}_{+}(t) } = \begin{pmatrix}
1-\rho_{\rm ad}(t) \\
\rho_{\rm ad}(t)
\end{pmatrix}, \kket{\lambda^{0}_{-}(t) } =\begin{pmatrix}
 \rho_{\rm ad}(t) \\
- \rho_{\rm ad}(t)
\end{pmatrix},& \\
&\bbra{l^{0}_{+}(t)} = \bbra{1},\ \ \ \bbra{l^0_-(t)} = \begin{pmatrix}
1, & - \frac{a_4(t)}{a_1(t)}\end{pmatrix}& \label{0lef}
\end{eqnarray}
where $\lambda(t) \equiv a_1(t)+a_4(t)$, and 
\begin{eqnarray}
\rho_{\rm ad}(t) \equiv \frac{a_1(t)}{\lambda(t)} = \frac{a_1(t)}{ a_1(t)+a_4(t)}.
\end{eqnarray} 
To derive Eqs.\  (\ref{0lam})--(\ref{0lef}) we have used the trivial relations $a_2^0(t)=-a_4(t)$ and $a_3^0(t)=-a_1(t)$. 
By solving Eq.\ (\ref{QME3}) under the condition $\chi =0$, one of the diagonal components of the density matrix is given by 
\begin{eqnarray}\label{rhot}
\rho_{11}(t) = \rho_{11}(0) e^{\int_0^td\tau\lambda(\tau)}- \int_0^td\tau a_1(\tau)e^{\int_{\tau}^td\tau'\lambda(\tau')},
\end{eqnarray}
where $\rho_{11}(t)$ represents $\bra{1}\rho(0,t)\ket{1}$. 
\begin{figure}[htbp]
\begin{minipage}{0.5\hsize}
  \includegraphics[scale =0.4]{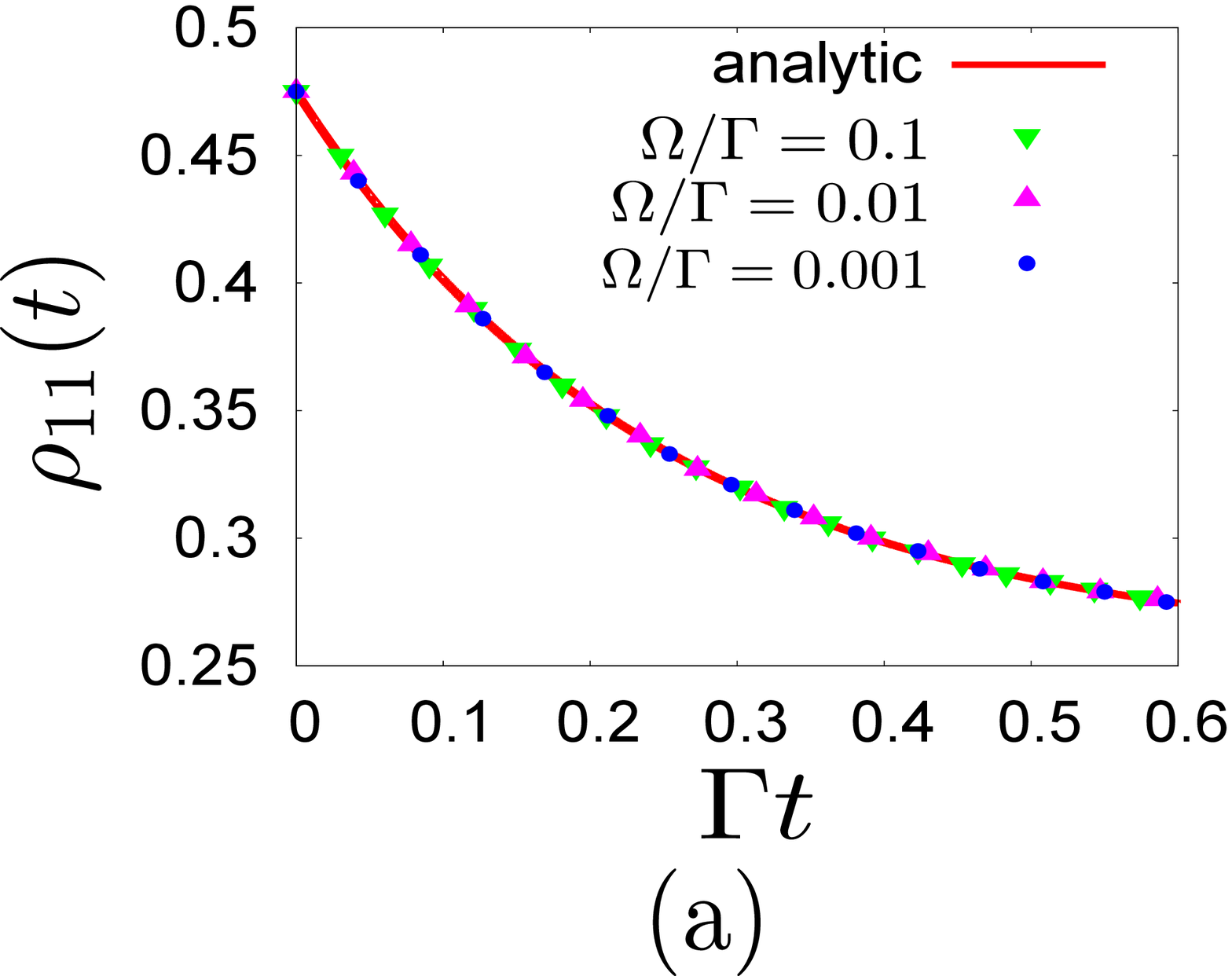}
\end{minipage}
\begin{minipage}{0.5\hsize}
  \includegraphics[scale =0.4]{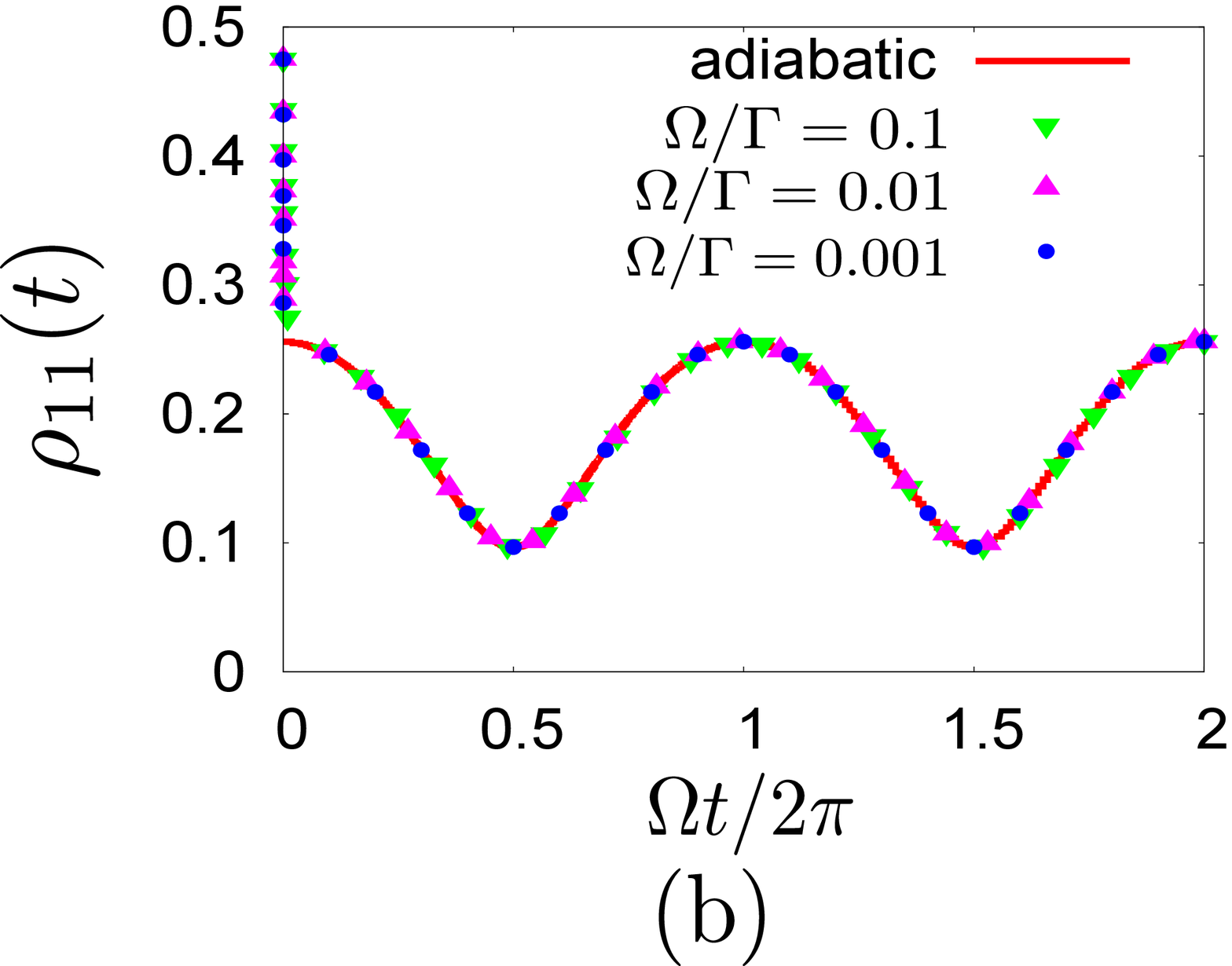}
\end{minipage}
 \caption{The time evolution of $\rho_{11}(t)$ at $(k_{\rm B}T_{0}/\hbar\omega_0)^{-1}=1.5$, $(k_{\rm B}T_{A}/\hbar\omega_0)^{-1}=3$, $\Gamma = 0.001\omega_0 $ and $\rho_{11}(0)=0.475$. (a) The initial relaxation of $\rho_{11}(t)$ against $\Gamma t$ for $\Omega/\Gamma \leq 0.1$. 
All numerical data are collapsed on the red line given by Eq.(\ref{fast}). (b) The long time behavior of $\rho_{11}(t)$ plotted against $\Omega t/2\pi$, where the red line represents the adiabatic form $\rho_{\rm ad}(t)$. } 
 \label{fig1}
\end{figure}

The time evolution of $\rho_{11}(t)$ is shown in Fig.\ref{fig1}.
For $\Gamma t \leq 1$, $\rho_{11}(t)$ can be approximated by
\begin{eqnarray}
\rho_{11}(t) \simeq \rho_{\rm ad}(t) + \{ \rho_{11}(0) - \rho_{\rm ad}(0) \}e^{\lambda(0)t}. \label{fast}
\end{eqnarray}

On the other hand, for $\Gamma t \gg 1$, $\rho_{11}(t)$ is asymptotically given by (see Appendix E)
\begin{eqnarray}\label{rhoasy}
\rho_{11}(\theta) \simeq \rho_{\rm ad}(\theta) + A_0(\theta) \frac{\Omega}{\Gamma} + A_1(\theta) \frac{\Omega^2}{\Gamma^2} + A_2(\theta)\frac{\Omega^3}{\Gamma^3}+O\left( \frac{\Omega^4}{\Gamma^4} \right),
\end{eqnarray}
where $\theta = \Omega t$ and 
\begin{eqnarray}
A_0(\theta) &=& -\frac{\rho_{\rm ad}'(\theta)}{2(1+n_L(\theta)+n_R(\theta))}, \\
A_1(\theta) &=&  \frac{\rho_{\rm ad}^{(2)}(\theta)}{4(1+n_L(\theta)+n_R(\theta))^2}  -\frac{(\rho_{\rm ad}'(\theta))^2}{2(1+n_L(\theta)+n_R(\theta))}, \\
A_2(\theta) &=&-\frac{\rho_{\rm ad}^{(3)}(\theta)}{8(1+n_L(\theta)+n_R(\theta))^3}  +\frac{\rho_{\rm ad}^{(2)}(\theta)\rho_{\rm ad}'(\theta)}{(1+n_L(\theta)+n_R(\theta))^2} -\frac{(\rho_{\rm ad}'(\theta))^3}{2(1+n_L(\theta)+n_R(\theta))}, \nonumber \\ \label{rhoasyco} 
\end{eqnarray}
where $\rho'_{\rm ad}(\theta)=d\rho_{\rm ad}(\theta)/d\theta$. In the adiabatic limit $\Omega/\Gamma \rightarrow 0$, $\rho_{11}(t)$ is reduced to $\rho_{\rm ad} (t)$ for $t \gg 1/\Gamma$.

\begin{figure}[h]
\centering
  \includegraphics[scale =0.35]{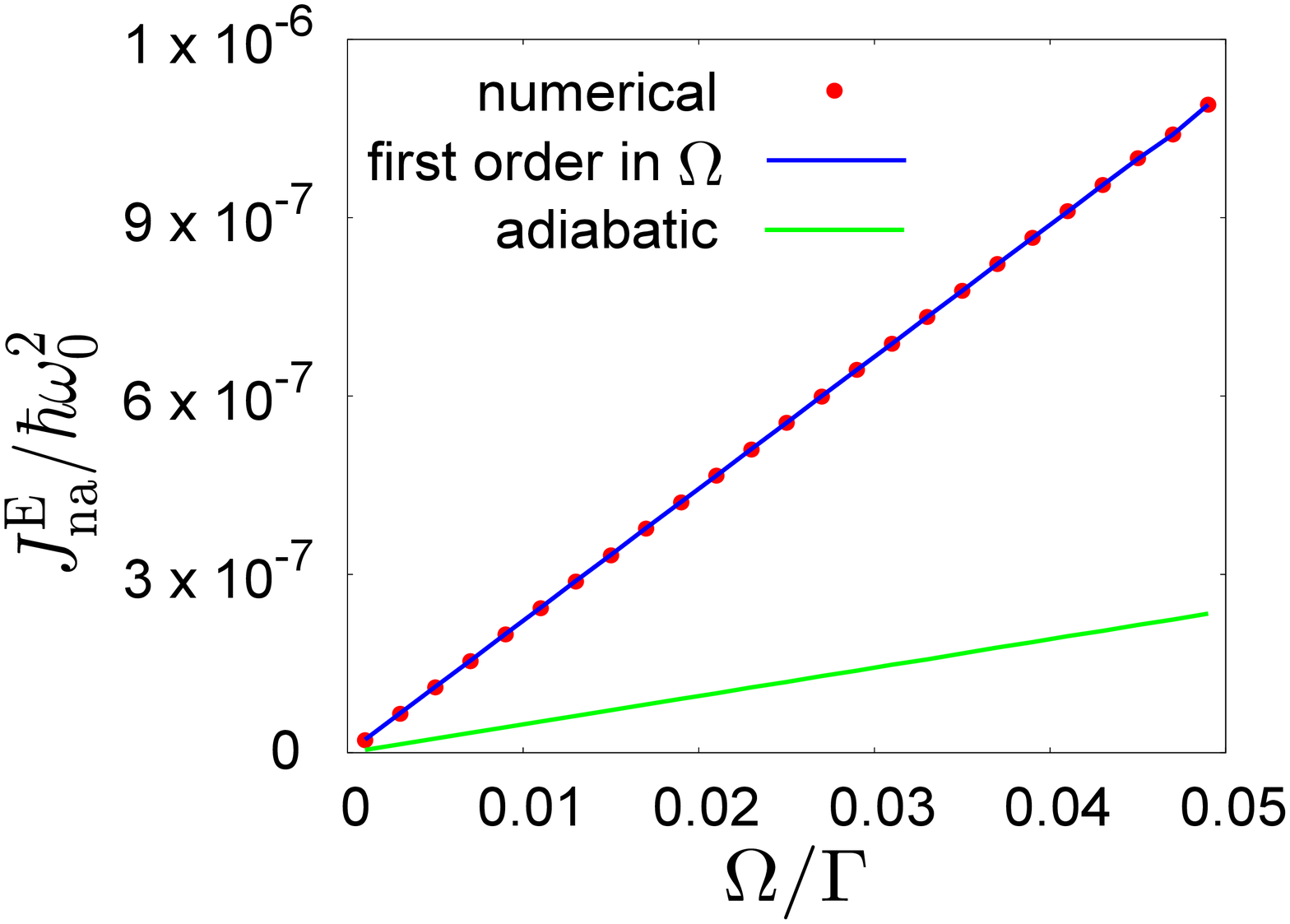}
 \caption{The plot of the pumping current against $\Omega/\Gamma$. The set of parameters is equivalent to that used in Fig.\ref{fig1}. The difference between the analytic result obtained from the first order of the asymptotic expansion of $J^{\rm E}_{\rm na}$ and the numerical result represented by the red dots is invisible.} 
 \label{fig2}
\end{figure}

Because we consider a system that has symmetric junctions between the system and the environments under the no-average bias, it is easy to show that $\expec{\Delta q_t}^{\rm d}$ is zero (see (\ref{Rendynsym})).
 We, thus, plot the non-adiabatic pumping current 
\begin{eqnarray}\label{exccurrent}
J^{\rm E}_{\rm na} = -\frac{1}{\tau_p} \int_{0}^{\tau_p}dt \bbra{l_+'(\bm{\beta}(t))}\df{}{t}\kket{\rho(0,t)} 
\end{eqnarray}
which is defined by $\expec{\Delta q_t}_{\rm na}^{\rm E}/\tau_p$, and the adiabatic one 
\begin{eqnarray}
J^{\rm g}_{\rm a} = -\frac{1}{\tau_p} \int_{0}^{\tau_p}dt \bbra{l_+'(\bm{\beta}(t))}\df{}{t}\kket{\lambda_+^0(t)} 
\end{eqnarray}
against the frequencies of modulation $\Omega$ with the numerical calculation and the asymptotic expansion (Fig.\ref{fig2}). It should be noted that large deviation between the adiabatic current and the obtained current mainly originates from the initial condition dependence. In other words, if we start the measurement of the current after $t\gg \lambda(0)^{-1}$, the  adiabatic approximation gives a reasonable result over wide range of $\Omega$.

\begin{figure}[h]
\centering
  \includegraphics[scale =0.35]{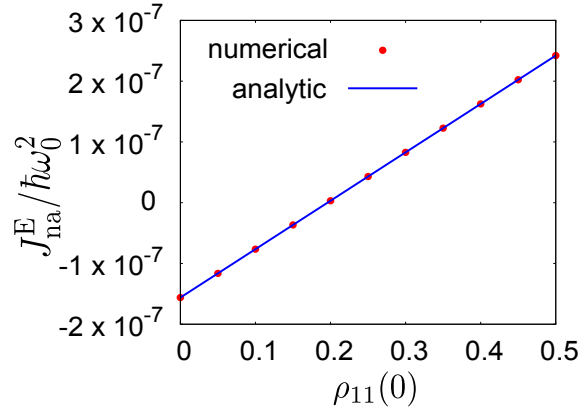}
 \caption{The plot of pumping $J^{\rm E}_{\rm na}$ vs the initial state $\rho_{11}(0)$ at $\Omega/\Gamma=0.01$, where ``analytic'' in the legend represents the expressions in Eqs.\ (\ref{pzen})--(\ref{pb2}). } 
 \label{fig3}
\end{figure}

As shown in Appendix E, the asymptotic expression of the pumping current is given by
\begin{eqnarray}
J^{E}_{\rm na}\simeq \Omega \left[ B_0 + B_1 \frac{\Omega}{\Gamma}+B_2\frac{\Omega^2}{\Gamma^3}+ O\left(\frac{\Omega^3}{\Gamma^3}\right)\right], \label{pzen}
\end{eqnarray}
where
\begin{eqnarray}
B_0 &=&- \frac{\hbar \omega_0}{4\pi} \int_0^{2\pi}d\theta \frac{1+2n_R(\theta)}{1+n_L(\theta)+n_R(\theta)}\rho_{\rm ad}'(\theta)\nonumber \\
& & +\frac{\hbar \omega_0}{4\pi}\frac{1+2n_R(0)}{1+n_L(0)+n_R(0)}(\rho_{11}(0) - \rho_{\rm ad}(0)), \label{b0} \\
B_1&=&  \frac{\hbar \omega_0}{8\pi}\int_0^{2\pi}d\theta \frac{1+2n_R(\theta)}{1+n_L(\theta)+n_R(\theta)} \left( \frac{\rho_{\rm ad}^{(2)}(\theta)}{1+n_L(\theta)+n_R(\theta)}-2(\rho_{\rm ad}'(\theta))^2\right) \nonumber \\
& & +\frac{\hbar \omega_0}{4\pi}\left[\frac{1+2n_R(0)}{1+n_L(0)+n_R(0)}\rho_{\rm ad}'(0) \left( \frac{1}{2}-\rho_{11}(0)\right)+ \frac{n_R'(0)(\rho_{11}(0) - \rho_{\rm ad}(0))}{(1+n_L(0)+n_R(0))^2}\right]. \label{b1} \\
B_2&=&  -\frac{\hbar \omega_0}{16\pi}\int_0^{2\pi}d\theta \frac{1+2n_R(\theta)}{1+n_L(\theta)+n_R(\theta)} \left(  \frac{\rho_{\rm ad}^{(3)}(\theta)}{(1+n_L(\theta)+n_R(\theta))^2}-\frac{8\rho_{\rm ad}^{(2)}(\theta)\rho_{\rm ad}'(\theta)}{1+n_L(\theta)+n_R(\theta)}+4(\rho_{\rm ad}'(\theta))^3\right) \nonumber \\
& & +\frac{\hbar \omega_0}{4\pi}\left[\frac{(1+2n_R(0))\rho_{\rm ad}''(0)}{(1+n_L(0)+n_R(0))^2} \left( \rho_{\rm ad}(0) -\frac{\rho_{11}(0)}{2}-\frac{1}{4}\right) +\frac{(1+2n_R(0))(\rho_{\rm ad}'(0))^2(\rho_{11}(0) - \rho_{\rm ad}(0))}{1+n_L(0)+n_R(0)} \right. \nonumber \\
& &+\left. \frac{n_R'(0)\rho_{\rm ad}'(0) }{(1+n_L(0)+n_R(0))^2}\left( 2\rho_{\rm ad}(0) -3\rho_{11}(0)+\frac{1}{2}\right)+\frac{n_R''(0)(\rho_{11}(0) - \rho_{\rm ad}(0))}{(1+n_L(0)+n_R(0))^3}\right]. \label{pb2} 
\end{eqnarray}
Thus, $J_{\rm na}^{\rm E}$ in the lowest order in $\Omega$ is reduced to the adiabatic pumping $J^{\rm g}_{\rm a}$ if we begin with $\rho_{11}(0) = \rho_{\rm ad}(0)$. If we begin with $\rho_{11}(0) \neq \rho_{\rm ad}(0)$, however, the expression of the adiabatic current does not give the correct result for the pumping current. 
To verify the results in Eqs.\ (\ref{pzen})--(\ref{pb2}), we explicitly plot how the pumping current depends on the initial condition(Fig.\ref{fig3}), where the analytic result (solid line) perfectly reproduces the numerical results. 

\begin{figure}[h]
\begin{minipage}{0.5\hsize}
  \includegraphics[scale =0.34]{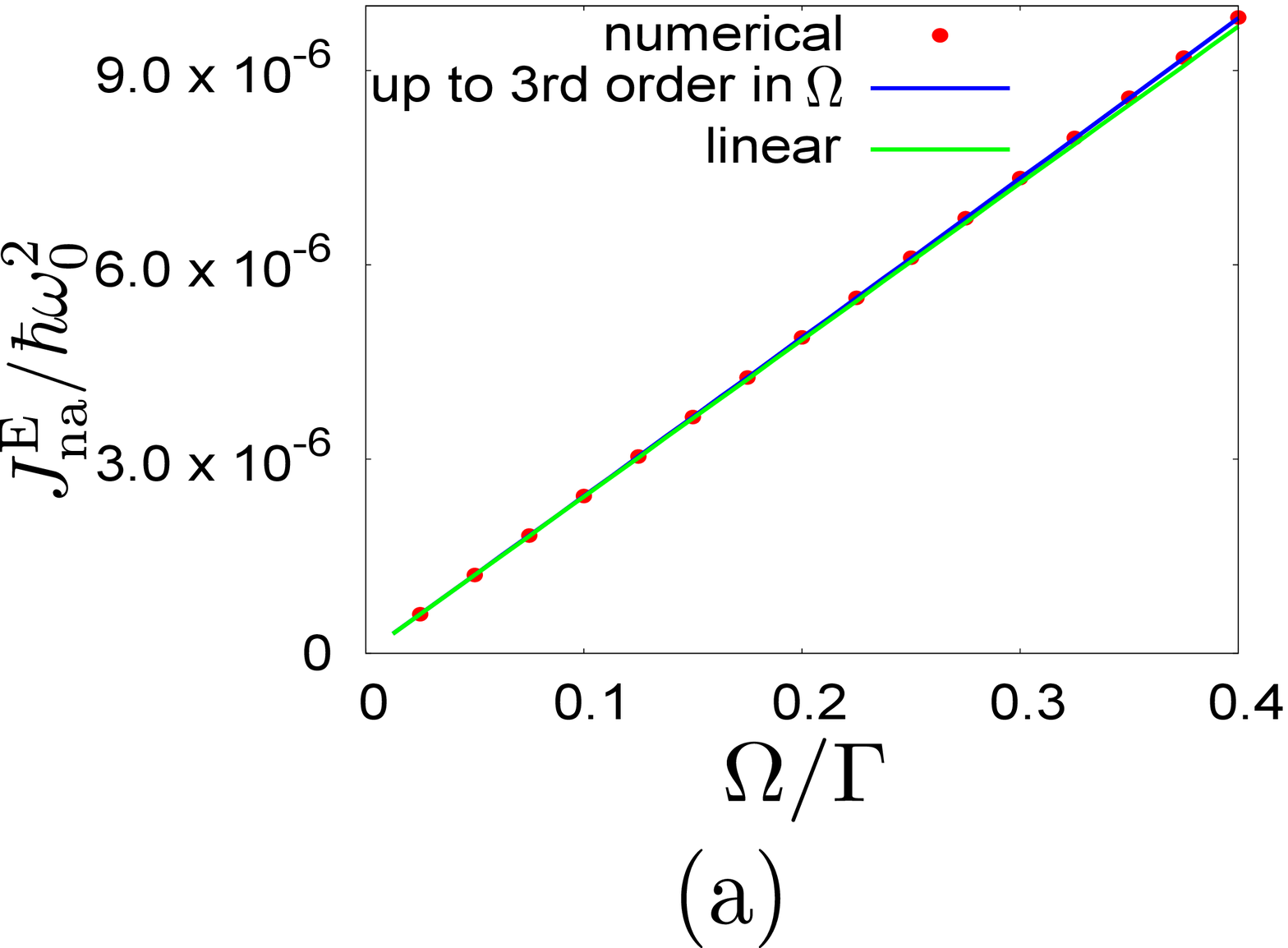}
\end{minipage}
\begin{minipage}{0.5\hsize}
  \includegraphics[scale =0.36]{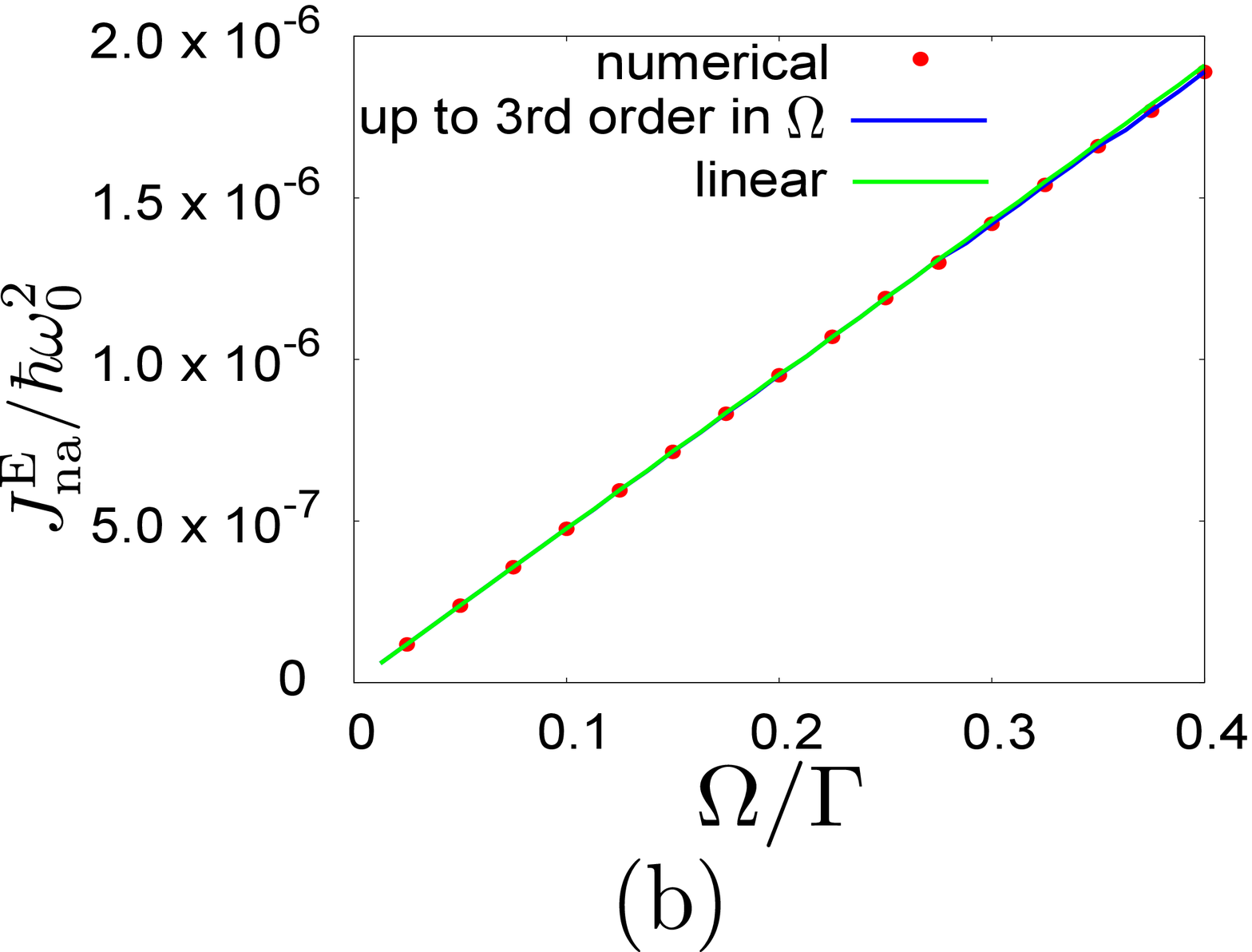}
\end{minipage}
 \caption{The plot of the pumping current $J^{\rm E}_{\rm na}$ and corresponding analytic calculations in Eqs.\ (\ref{pzen})--(\ref{pb2}) up to the first and third orders in $\Omega$ for (a) $\rho_{11}(0)=0.5$ and (b) $\rho_{11}(0)=\rho_{\rm ad}(0)$. The line with the legend ``linear'' represents the analytic expression up to $O(\Omega)$ in Eqs.\ (\ref{pzen}) and (\ref{b0}). The set of parameters is equivalent to that used in Fig.\ref{fig1}} 
 \label{fig4}
\end{figure}

\begin{figure}[h]
\begin{minipage}{0.5\hsize}
  \includegraphics[scale =0.37]{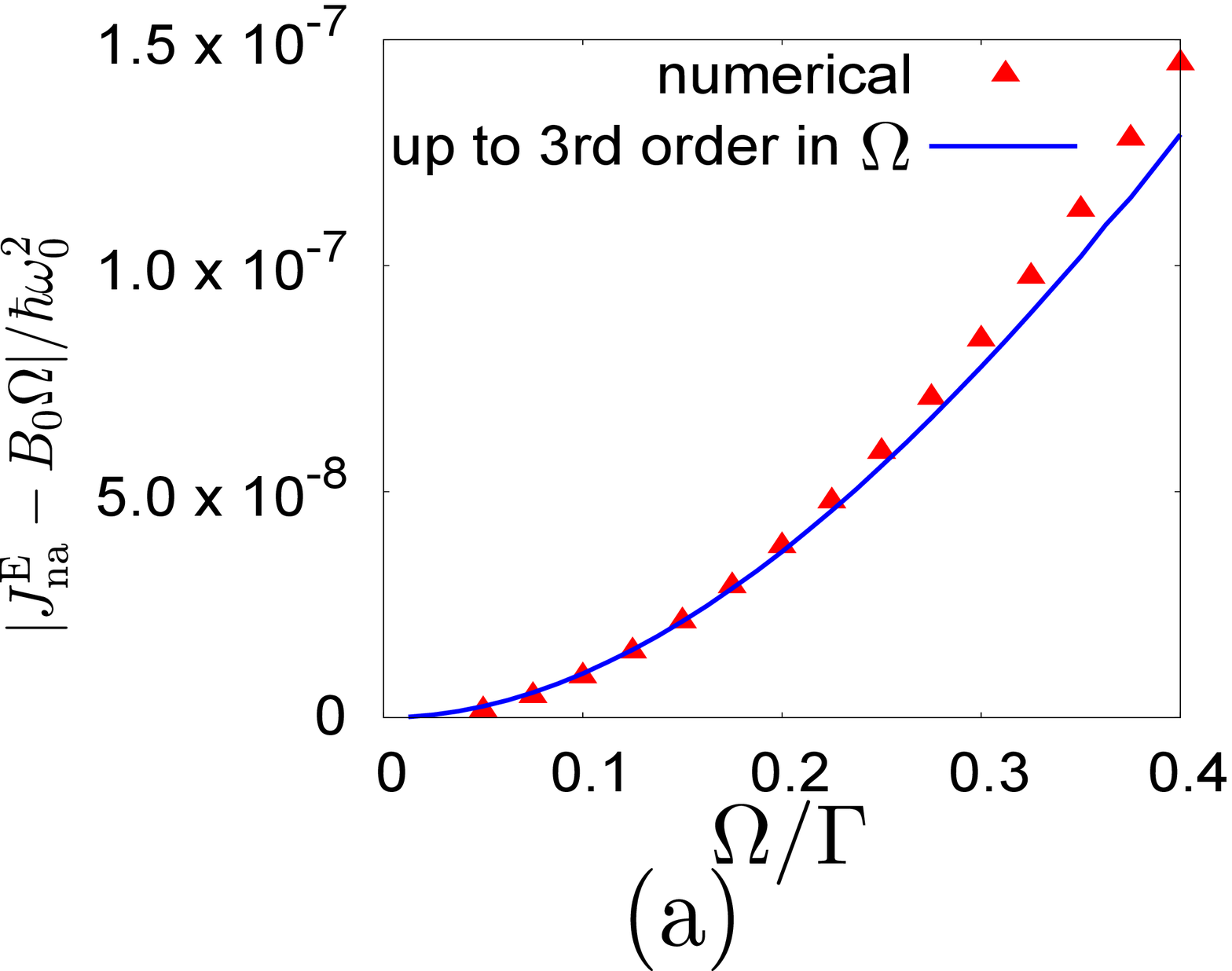}
\end{minipage}
\begin{minipage}{0.5\hsize}
  \includegraphics[scale =0.39]{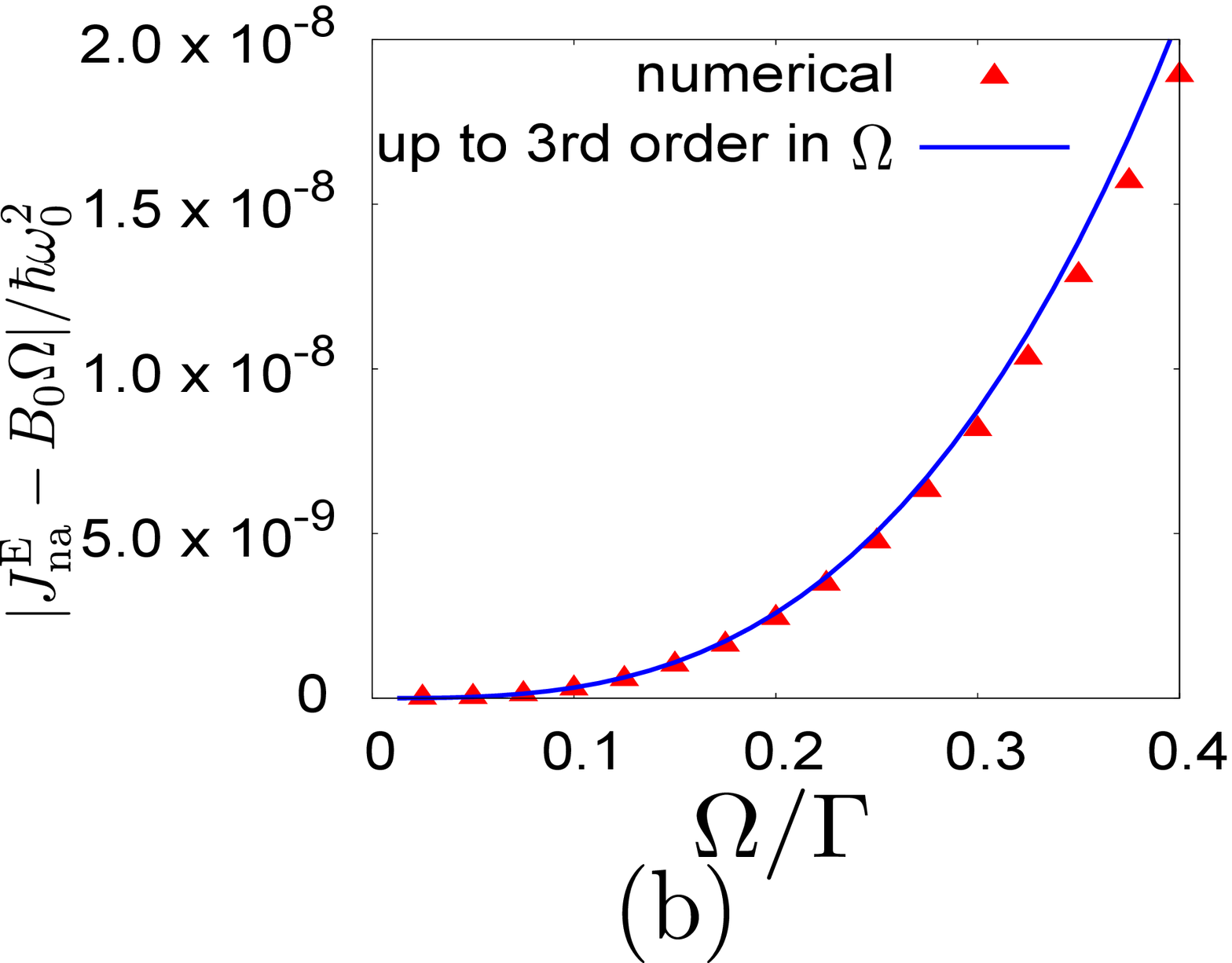}
\end{minipage}
 \caption{The non-adiabatic pumping current obtained from the subtraction of the first-order term $B_0\Omega$ to clarify the term up to the third order for (a) $\rho_{11}(0) =0.5$ and (b) $\rho_{11}(0)=\rho_{\rm ad}(0)$. We use identical parameters to those used in Fig.\ref{fig1}. The non-adiabatic  pumping current obtained from the analytic calculation (blue line) asymptotically reproduces the numerical result (red dots) in the small region of $\Omega/\Gamma$. }
 \label{fig5}
\end{figure}

In the case $\rho_{11}=0.5$, we plot the result of the pumping current in Fig.\ref{fig4}(a). To remove the initial condition dependence, we also plot the result of the pumping current for the initial measurement starting from $t\gg \lambda(0)^{-1}$ in Fig.\ref{fig4}(b), where we start the measurement from $\tau_p$, the initial condition  of which corresponds to $\rho_{11}(0)=\rho_{\rm ad}(0)$. 
It is clear that the adiabatic current gives a reasonable result for $\Omega/\Gamma<0.2$, but there exists a little systematic deviation between the linear or adiabatic result and the numerical result for $\Omega/\Gamma>0.2$. To clarify the non-adiabatic contribution up to $\Omega^3$, we plot $J-B_0\Omega$ (see Figs.\ref{fig5}(a) and (b)). It is obvious that our analytic non-adiabatic expression in Eqs.\ (\ref{pzen})--(\ref{pb2}) gives a reasonable result even if the linear expression in $J_{\rm na}^{\rm E}$ is no longer valid.

\subsection{Extended fluctuation theorem}
In this subsection, we discuss whether the fluctuation theorem for the heat currents exists in our system under the existence of the dc bias, i.e. $\Delta \beta \equiv \beta_R-\beta_L$. 
In this sense, the set-up of the heat fluctuation theorem in this subsection differs from the case without dc bias, discussed elsewhere.

First, we consider the case without temporal temperatures change, i.e.  $\beta_{\nu}(t)=\beta_{\nu}$. 
Then, we readily obtain $J^{\rm g}_{\rm a}=J^{\rm E}_{\rm na}=0$ because $\rho(\chi,t)$ decays with negative eigenvalues in Eq.(\ref{lambda}). 
Therefore, the steady cumulant-generating function is reduced to $S(\chi)\equiv \lim_{\tau \rightarrow \infty}\frac{1}{\tau}S(\chi,\tau)=\lambda_+^{\chi}$, and the 
product of $a_2^{\chi}$ and $a_3^{\chi}$ satisfies the Gollavotti--Cohen (GC) symmetry\cite{Lebowitz}:  
\begin{eqnarray}\label{GCs}
a_2^{\chi}a_3^{\chi} = a_2^{-\chi+i\alpha}a_3^{-\chi+i\alpha},
\end{eqnarray}
where $\alpha = \ln{\frac{n_L(1+n_R)}{n_R(1+n_L)}}/\hbar \omega_0 = \Delta \beta$. If the cumulant-generating function satisfies GC symmetry, the steady fluctuation theorem holds:
\begin{eqnarray}
\lim_{\tau \rightarrow \infty} \frac{1}{\tau} \ln{\frac{P(\Delta q_{\tau})}{P(-\Delta q_{\tau})}} = \Delta \beta \frac{\Delta q_{\tau}}{\tau}.
\end{eqnarray}

Nevertheless, the simple GC symmetry for the cumulant-generating function no longer holds when the temperature varies with time even in the adiabatic limit because the constant $\alpha$ is replaced by $\alpha (t)$ and the contribution of the geometrical term exists. 
This may give the basis of the violation of the heat fluctuation theorem in Ref.\cite{Ren}, and the geometrical entropies introduced in Refs.\cite{Sagawa, Yuge13}. 
In the case of the periodic change of temperatures, there is room to choose a variable of large deviation to obtain the correlation of the fluctuation theorem.

In the case $\tau_p\omega_0 \rightarrow \infty$, the fluctuation theorem for the current $\xi(t)= d\Delta q_t/dt$ is given as follows (see Appendix F.1)
\begin{eqnarray}\label{tft}
\frac{1}{\tau_p\omega_0}\ln \frac{P(\xi)}{P(-\xi)} = \overline {\Delta \beta(t) \xi(t)} - \frac{1}{2\tau_p\omega_0} \ln\frac{\{ \overline{ v^{\chi}(\theta) }^3 \partial^2 \overline{\lambda_+^{\chi}(t)}/\partial \chi^2\}|_{\chi=\chi^*(\overline{\xi})}}{\{ \overline{v^{\chi}(\theta) }^3 \partial^2 \overline{\lambda_+^{\chi}(t)}/\partial \chi^2\}|_{\chi=\chi^*(-\overline{\xi})}} + O((\tau_p\omega_0)^{-2}),
\end{eqnarray}
where $\overline{A}$ expresses the time average of an arbitrary valuable $A$ during the time interval $\tau_p$, $\chi^*(\xi)$ satisfies $i\overline{\xi} = \partial \overline{\lambda_+^{\chi^*}(t)}/\partial \chi^*$ and $v^{\chi}(\theta)$ is a dimensionless geometrical term $v^{\chi}(\theta) = \bbra{l_+^{\chi}(\theta)}d\kket{\lambda_+^{\chi}(\theta)}/d\theta$ since we consider $\tau_p\omega_0 \rightarrow \infty$. We note that the formula (\ref{tft}) can be applied even for the case $\overline{\Delta \beta(t)} = 0$.

On the other hand, for $N = \Delta q/\hbar\omega_0 \rightarrow \infty$, the fluctuation thorem for the transferred energy is given by
\begin{eqnarray}\label{sft}
\ln \frac{P_F(+N\hbar \omega_0)}{P_B(-N\hbar \omega_0)} &=& N\hbar \omega_0  \overline{\Delta \beta} + \hbar \omega_0 \frac{\overline{ \Delta \beta} -\Delta \beta(0)}{2} \nonumber \\
& &+\frac{\hbar \omega_0\pi}{4N}\left\{  u_L(0)-u_R(0) \right\} + O\left( N^{-2} \right), 
\end{eqnarray}
where $u_{\nu}(\theta) = (5+2n_{\nu}(\theta))d\beta_{\nu}(\theta)/d\theta$ and $P_B$ is the time-reversal distribution against $P_F$ as shown in Appendix F.2.

Numerical verification of the extended fluctuation theorems will be reported elsewhere. Nevertheless, we believe that the extended fluctuation theorem in Eq.\ (\ref{tft}) is, at least, universal for the slowly modulated case. Indeed, the derivation of Eq.\ (\ref{tft}) does not contain any specific feature of the spin-boson system.

\section{Discussion and conclusion}
 We have successfully extended the theory of adiabatic pumping to non-adiabatic pumping for finite speed modulations within the framework of the Markovian quantum master equation.
 We have applied our formulation to the spin-boson system and found that (i) the pumping current strongly depends on the initial condition, and (ii) the contribution of the non-adiabatic pumping current is relevant for relatively large $\Omega/\Gamma$ if the contribution of the initial relaxation is eliminated.
(iii) The contribution of the non-adiabatic effect is analytically reproducible in terms of the technique of the asymptotic expansion. 
(iv) The extended fluctuation theorems for slowly modulated temperatures and large transferred energies are derived.

Our master equation in a weak coupling limit does not have any contribution from the off-diagonal elements of the density matrix. Therefore, our master equation is reduced to the classical rate equation\cite{Ren}. To extract the off-diagonal contributions, we may consider a strong coupling regime or more complicated model such as a three-level system. This will be our future work.

Although our system is equivalent to that analyzed in Ref.\cite{Uchiyama}, there are various differences in the analysis between two papers. Indeed, Ref.\cite{Uchiyama} uses discretized time evolution under the finite interval $\tau_p/41$ when the master equation is solved, while we have obtained both numerical and analytic solutions under continuous time evolution. Moreover, we have explicitly obtained the analytic form for the pumping current whose validity is quantitatively verified through comparison with the numerical calculation. Furthermore, the discussion on the extended fluctuation theorems in Sect. 3.3 is completely new. Therefore, we believe that there exist several merits for the publication of our paper besides Ref.\cite{Uchiyama}. 
 
We have derived an analytic expression for the non-adiabatic pumping current $J^{\rm E}_{\rm na}$ corresponding to the geometrical phase in the adiabatic limit. We should note, however, that $J^{\rm E}_{\rm na}$ is no longer geometric quantity as in the adiabatic case because the curvature depends on time. Such a time-dependent quantity may be interpreted by the Aharonov-Anandan phase method\cite{Aharonov}.

We have also derived the fluctuation theorem with the temporal change of parameters in the case of $\tau_p\omega_0 \rightarrow \infty$ or $N \rightarrow \infty$. This is a natural extension of the steady fluctuation theorem to the time-dependent fluctuation theorems.

We expect Floquet theory to be applicable to our system because we study periodic modulations to the system. In future work we will compare our analysis with that based on Floquet theory.

 We only analyze the case of continuous modulation of the temperature (\ref{T(t)}) under the assumption that both environments are always in equilibrium. It is straightforward to apply our formulation to fermion systems such as the impurity Anderson model\cite{Yoshii}.

It should be noted that our theory cannot be applied to either discontinuous changes in temperature or fast modulation. 
Although it is possible to apply our formulation to non-Markovian processes, 
we are suspicious of whether the analysis within this framework is meaningful, because non-Markovian processes may affect the state of the environments.
We also note that the distinction between two current terms is no longer valid for non-Markovian processes, 
because $\lambda(t)$ sometimes takes positive values. 
We will discuss the non-Markovian pumping process elsewhere.

\section*{Acknowledgements}
The authors thank C. Uchiyama for her collaboration in the early stage of this work, 
J. Ohkubo for his helpful advice and R. Yoshii, S. Nakajima, T. Sagawa and Y. Watanabe for
valuable discussions.
This work is partially supported by a Grant-in-Aid of MEXT (Grant No. 25287098).

\appendix
\section{Cumulant-generating function}

 In this appendix, we briefly summarize the relationship between FCS and the cumulant-generating function.
 Let us perform a projection measurement of $Q$ at 0 and $t$, where their measured values are set to be $q_0$ and $q_t$, respectively.
Here, we assume that $Q$ satisfies $[Q, \rho_{\rm tot}(0)]=0$, where $\rho_{\rm tot}(t)$ is the total density matrix at time $t$ without the counting field. 
The probability of measuring $q_0$ and $q_t$ is given by
\begin{eqnarray}
P[q_t,q_0]=\tr\left( P_{q_t} U(t,0) P_{q_0}\rho_{\rm tot}(0) P_{q_0}U(t,0)^{\dagger}P_{q_t} \right),
\end{eqnarray}
where $P_{q_0}$ is the projection operator onto the eigenstates corresponding to the eigenvalue $q_0$ and $U(t,0)$ is the unitary time evolution operator of the total system, which is defined by 
\begin{eqnarray}
\df{}{t}U(t,0) = -\frac{i}{\hbar}H_{\rm tot}(t) U(t,0)
\end{eqnarray}
satisfying $U(t,t)=1$, and $U(t,0)^\dagger = U(0,t)$ is the adjoint matrix of $U(t,0)$, where $H_{\rm tot}(t)$ is the total Hamiltonian.
Thus, the probability of the current $\Delta q_t$ at $t$ is given by
\begin{eqnarray}
P(\Delta q_t)=\sum_{q_0,q_t}\delta( q_t -q_0 -\Delta q_t) P[q_t,q_0].
\end{eqnarray}
Let us introduce the characteristic function as
\begin{equation}\label{def_generating_fn}
G(\chi,t)\equiv 
\int d \Delta q_t e^{i\chi \Delta q_t}P(\Delta q_t)
=\sum_{q_0,q_t}e^{i\chi (q_t-q_0)} P[q_t,q_0].
\end{equation}
From the identities $P_{q_0}^2=P_{q_0}, q_0P_{q_0}=QP_{q_0}, \sum_{q_0}P_{q_0}=1$, we can rewrite (\ref{def_generating_fn}) as
\begin{eqnarray}
G(\chi,t)=\tr\left(U_{\chi/2}(t,0) \rho_{\rm tot}(0)U_{-\chi/2}(t,0)^{\dagger} \right), \label{gf}
\end{eqnarray}
where $U_{\chi/2}(t,0)$ is defined by 
\begin{eqnarray}
\df{}{t}U_{\chi}(t,0) = -\frac{i}{\hbar}H_{\rm tot,\chi}(t) U_{\chi}(t,0)
\end{eqnarray}
and $H_{\rm tot,\chi} \equiv e^{i\chi Q}H_{\rm tot}(t)e^{-i\chi Q}$.
Equation (\ref{gf}) automatically satisfies $G(\chi,0)=1$ because of the relation $U_{\chi}(t,0) \rightarrow 1$ in the limit $t\rightarrow +0$. 
Thus, the cumulant-generating function 
\begin{equation}\label{cumulant_def}
S(\chi,t)\equiv \ln{G(\chi,t)}
\end{equation}
satisfies $S(\chi,0)=0$. 
Hence, all of the cumulants at $t=0$ satisfy 
\begin{eqnarray}
\expec{\Delta q^n}_c = 0 .
\end{eqnarray}
We introduce the total modulated density matrix
\begin{eqnarray}
\rho_{\rm tot}(\chi,t) \equiv U_{\chi/2}(t,0) \rho_{\rm tot}(0)U_{-\chi/2}(t,0)^{\dagger} \label{modrho}
\end{eqnarray}
and the modulated density matrix of the system is $\rho(\chi,t) \equiv \tr_{\rm E}\rho_{\rm tot}(\chi,t)$.

Let us rewrite $S(\chi,t)$ as 
\begin{eqnarray}
S(\chi,t)= \ln \tr_{\rm S} \rho(\chi,t). \label{rhotog}
\end{eqnarray}
Note that the argument presented here is still valid even for non-Markovian case.

\section{Derivation of the quantum master equation with parameters modulation}
In this appendix, we derive the FCS quantum master equation under the modulation of parameters though the derivation of the master equation without modulation is well known\cite{Breuer,Esposito}. The total Hamiltonian is given by
\begin{eqnarray}
H_{\rm tot} &=& H_0 + gH_{\rm SE}, \\
H_0 &=& H_{\rm S} + H_{\rm E},
\end{eqnarray}
where $H_{\rm S}$ is the Hamiltonian of the target system,  $H_{\rm E}$ is the Hamiltonian of environments, and $H_{\rm SE}$ is the interaction between the system and environments characterized by the coupling constant $g$. 

The total system with the full counting statistics is expressed by the modified von Neumann equation from Eq.\ (\ref{modrho})
\begin{eqnarray}
\df{}{t}\rho_{\rm tot}(\chi,t) =-i \mathcal L^{\chi} \rho_{\rm tot}(\chi,t), \label{mvneq}
\end{eqnarray}
where the modified Liouvillian $\mathcal L^{\chi}(t)$ is defined by
\begin{eqnarray}
 \mathcal L^{\chi} &=&  \mathcal L_0 + g \mathcal L^{\chi}_{\rm SE}, \\
\mathcal L_0 \rho&=&\frac{1}{\hbar}[H_0, \rho] ,\\
\mathcal L^{\chi}_{\rm SE} \rho &=& \frac{1}{\hbar}[H_{\rm SE}, \rho] _{\chi},
\end{eqnarray}
and $[H_{\rm SE}, \rho] _{\chi} = H_{\rm SE,\chi}\rho- \rho H_{\rm SE,-\chi}$.

The formal solution of Eq.\ (\ref{mvneq}) can be written as
\begin{eqnarray}
\rho_{\rm tot}(\chi,t) = e^{ -i\mathcal L^{\chi}t} \rho_{\rm tot}(\chi,0). \label{formals}
\end{eqnarray}

To trace out the degree of freedom of environments, we introduce the Nakajima--Zwanzig projection operator
\begin{eqnarray}
\mathcal P(t)\rho &=& \rho^{\rm eq}_{\rm E}({\bm \pi}_t) \tr_{\rm E}\rho, \\
\mathcal Q(t) &=& 1 -\mathcal P(t),
\end{eqnarray}
where $\rho^{\rm eq}_{\rm E}({\bm \pi}_t)$ is the equilibrium operator of environments and under the modulation of parameters ${\bm \pi}_t$, such as temperature and chemical potential. 
We assume that environments are always at equilibrium during the modulation of parameters ${\bm \pi}_t$. Even if we adopt such a simplification, the derivation of the master equation is non-trivial, because ${\bm \pi}_t$ depends on time. The projection operators satisfy $\mathcal P(t) \mathcal P(t') = \mathcal P(t)$, $\mathcal P(t) \mathcal Q(t') = 0$, $\mathcal P(t)\mathcal L_0= \mathcal L_0\mathcal P(t)$. From the relation $\tr_{\rm E} (\rho^{\rm eq}_{\rm E} H_{\rm SE}) = 0$, we, thus, obtain 
\begin{eqnarray}
&\mathcal P(t'') \mathcal L^{\chi}_{\rm SE}\mathcal P(t) = 0,& \\
&\mathcal Q(t'') \mathcal L^{\chi} \mathcal P(t)  = \mathcal L^{\chi}_{\rm SE} \mathcal P(t),&\label{intqp} \\
&\mathcal P(t'') \mathcal L^{\chi} \mathcal Q(t)  = \mathcal P(t'')\mathcal L^{\chi}_{\rm SE}. & \label{intpq}
\end{eqnarray}

Let us introduce the projected time evolution operator 
\begin{eqnarray}
&\mathcal X(t) \equiv \mathcal P(t) e^{ -i\mathcal L^{\chi}t},& \\
&\mathcal Y(t) \equiv \mathcal Q(t) e^{ -i\mathcal L^{\chi}t}.&
\end{eqnarray}
The time evolutions of $\mathcal X$ and $\mathcal Y$ can be written as
\begin{eqnarray}
\df{}{t}\mathcal X(t) = \left\{ \mathcal P(t) (-i\mathcal L^{\chi}) + \df{\mathcal P(t)}{t} \right\}\mathcal X(t)+\mathcal P(t) (-i\mathcal L^{\chi} )\mathcal Y(t), \label{tevox}\\
\df{}{t}\mathcal Y(t) = \left\{ \mathcal Q(t) (-i\mathcal L^{\chi} ) - \df{\mathcal P(t)}{t} \right\}\mathcal X(t)+\mathcal Q(t) (-i\mathcal L^{\chi} )\mathcal Y(t). \label{tevoy}
\end{eqnarray}

The formal solution of Eq.\ (\ref{tevoy}) is given by
\begin{eqnarray}
\mathcal Y(t) = \int_0^t d\tau \widetilde{ \mathcal U}(t,\tau) \left\{ \mathcal Q(\tau) (-i\mathcal L^{\chi} ) - \df{\mathcal P(\tau)}{\tau} \right\} \mathcal X (\tau) + \widetilde{ \mathcal U}(t,0) \mathcal Q(0), \label{yformal}
\end{eqnarray}
where we have introduced
\begin{eqnarray}
 \widetilde{ \mathcal U}(t,t') = \mathcal Q(t')T_{\rightarrow} \exp{ \left( \int_{t'}^t d\tau \mathcal Q (\tau) (-i\mathcal L^{\chi}) \right)}. \label{tilu}
\end{eqnarray}

By using $\mathcal X(\tau) = \mathcal P(\tau) e^{ -i\mathcal L^{\chi}(\tau-t)}e^{ -i\mathcal L^{\chi}t} = \mathcal P(\tau) e^{ -i\mathcal L^{\chi}(\tau-t)}(\mathcal X(t) + \mathcal Y(t) )$ and
\begin{eqnarray}
 \mathcal S(t) \equiv \int_0^t d\tau \widetilde{ \mathcal U}(t,\tau) \left\{ \mathcal Q(\tau) (-i\mathcal L^{\chi} ) - \df{\mathcal P(\tau)}{\tau} \right\} \mathcal P(\tau) e^{ -i\mathcal L^{\chi}(\tau-t)},
\end{eqnarray}
Eq.\ (\ref{yformal}) can be rewritten as 
\begin{eqnarray}
\mathcal Y(t) = \{ 1 -\mathcal S(t) \}^{-1} \mathcal S(t) \mathcal X (t) +  \{ 1 -\mathcal S(t) \}^{-1} \widetilde{ \mathcal U}(t,0) \mathcal Q(0). \label{ysol}
\end{eqnarray}
Substituting Eq.\ (\ref{ysol}) into Eq.\ (\ref{tevox}), we obtain 
\begin{eqnarray}
\df{}{t}\mathcal X(t) &=& \left[ \mathcal P(t) (-i\mathcal L^{\chi} ) \{ 1 -\mathcal S(t) \}^{-1} + \df{\mathcal P(t)}{t} \right]\mathcal X(t) \nonumber \\
& &+\mathcal P(t) (-i\mathcal L^{\chi} ) \{ 1 -\mathcal S(t) \}^{-1} \widetilde{ \mathcal U}(t,0) \mathcal Q(0).\label{allox}
\end{eqnarray}

To perform the perturbation for the small coupling constant $g$, we rewrite $e^{-i\mathcal L^{\chi}t}$ and  Eqs.\ (\ref{tilu}) as 
\begin{eqnarray}
e^{ -i\mathcal L^{\chi}(t-t')} &=& e^{ -i\mathcal L_0(t-t')}\mathcal V(t,t'), \label{totud}\\
 \widetilde{ \mathcal U}(t,t') &=& e^{ -i\mathcal L_0(t-t')}\mathcal Q (t') \widetilde{\mathcal V}(t,t'),\label{tottilud}
\end{eqnarray}
where we have introduced
\begin{eqnarray}
\mathcal V(t,t') &=& T_{\rightarrow} \exp{ \left( \int_{t'}^t d\tau  (-ig\mathcal L_{\rm SE}^{\chi}(\tau)) \right)}, \label{fullv}\\
\widetilde{\mathcal V}(t,t') &=& T_{\rightarrow} \exp{ \left( \int_{t'}^t d\tau \mathcal Q (\tau)(-ig\mathcal L_{\rm SE}^{\chi}(\tau)) \right)}, \label{fulltv}
\end{eqnarray}
and $ \mathcal L_{\rm SE}^{\chi}(\tau) = e^{ i\mathcal L_0\tau}\mathcal L_{\rm SE}^{\chi}e^{ -i\mathcal L_0\tau} $.

In the small $g$ limit, Eqs.\ (\ref{totud}) and (\ref{tottilud}) are reduced to
\begin{eqnarray}
e^{ -i\mathcal L^{\chi}(\tau-t)} &\simeq& \left[1 - \int_{\tau}^t d\tau'  (-ig\mathcal L_{\rm SE}^{\chi}(\tau')) + O(g^2)\right]e^{ i\mathcal L_0(t-\tau)}, \\
\widetilde{ \mathcal U}(t,\tau) &\simeq&  e^{-i\mathcal L_0(t-\tau)}\mathcal Q(\tau)\left[1 + \int_{\tau}^t d\tau'  (-ig\mathcal L_{\rm SE}^{\chi}(\tau'))+ O(g^2 )\right].
\end{eqnarray}
Thus, $\mathcal S(t)$ reduces to
\begin{eqnarray}
 \mathcal S(t) \simeq \int_0^t d\tau  \mathcal Q(\tau) (-ig\mathcal L_{\rm SE}^{\chi}(\tau-t)) \mathcal P(\tau) + \mathcal M(t)+ O(g^2),
\end{eqnarray}
where we have used Eq.\ (\ref{intqp}) and defined $\mathcal M(t)$ as
\begin{eqnarray}
\mathcal M(t) \equiv \int_0^td\tau  e^{ -i\mathcal L_0 (t-\tau)}\left[\df{\mathcal P(\tau)}{\tau}\int_{\tau}^t d\tau'  (-ig\mathcal L_{\rm SE}^{\chi}(\tau'))-\int_{\tau}^t d\tau'  (-ig\mathcal L_{\rm SE}^{\chi}(\tau'))\df{\mathcal P(\tau)}{\tau} \right] e^{ i\mathcal L_0 (t-\tau)} \nonumber \\
\end{eqnarray}

From the relation $\{1-\mathcal S(t)\}^{-1} \simeq 1+\mathcal S(t)$ under the weak coupling limit, Eqs.\ (\ref{intpq}) and (\ref{allox}) can be rewritten as
\begin{eqnarray}
\df{}{t}\mathcal X(t) &\simeq& \left[ \mathcal P(t) (-i\mathcal L^{\chi} ) + \df{\mathcal P(t)}{t} \right]\mathcal X(t) \nonumber \\
& &+\int_0^t d\tau   \mathcal P(t) (-ig\mathcal L_{\rm SE}^{\chi} )  (-ig\mathcal L_{\rm SE}^{\chi}(\tau-t) ) \mathcal P(\tau)\mathcal X(t) \nonumber \\
& &+\mathcal P(t) (-i\mathcal L^{\chi})\mathcal M(t)\mathcal X(t) +\mathcal P(t) (-i\mathcal L^{\chi} ) \{ 1 -\mathcal S(t) \}^{-1} \widetilde{ \mathcal U}(t,0) \mathcal Q(0). \label{xevb}
\end{eqnarray}
When we operate Eq.\ (\ref{xevb}) on $\rho_{\rm tot}(\chi,0)$, we obtain the master equation
 \begin{eqnarray}
\df{}{t}\rho(\chi,t) &=& -\frac{i}{\hbar} [H_{\rm S}, \rho(\chi,t)] \nonumber \\
& &-\frac{g^2}{\hbar^2}\int_0^t d\tau  \tr_{\rm E} [H_{\rm SE}, [ e^{iH_0(\tau-t)/\hbar} H_{\rm SE}e^{-iH_0(\tau-t)/\hbar}, \rho^{\rm eq}_{\rm E}({\bm \pi}_{\tau})\rho(\chi,t) ]_{\chi}]_{\chi}\nonumber \\
& &+m(t) +I(t), \label{qmmas}
\end{eqnarray}
where we have introduced
 \begin{eqnarray}
m(t) = g^2\tr_B\left\{ \int_0^td\tau \mathcal L_{\rm SE}^{\chi} e^{-i\mathcal L_0(t-\tau)} \int_{\tau}^t d\tau' \mathcal L_{\rm SE}^{\chi}(\tau') \df{\mathcal P(\tau)}{\tau} e^{i\mathcal L_0(t-\tau) }\rho_{\rm tot}(\chi,t) \right\}
\end{eqnarray}
and the initial correlation term 
\begin{eqnarray}
I(t) =\tr_{\rm E}\{ (-i\mathcal L^{\chi}(t) ) \{ 1 -\mathcal S(t) \}^{-1} \widetilde{ \mathcal U}(t,0) \mathcal Q(0)\rho_{\rm tot}(\chi,0)\},
\end{eqnarray}
which vanishes if $\rho_{\rm tot}(\chi,0) =\rho(\chi,0)\rho^{\rm eq}_{\rm E}({\bm \pi}_0)$.

There exist several characteristic timescales in Eq.\ (\ref{qmmas}): the timescale $\tau_{\rm S}$ for the energy level of the system $H_{\rm S}$, the relaxation timescale $\tau_{\rm R}$ of the system, the correlation timescale $\tau_{\rm C}$ of the environments and the timescale $\Omega^{-1}$ of the modulation of parameters. $\tau_{\rm C}$ is the timescale that characterizes the symmetrized time correlation function $\tr_{\rm E}[\{B_i(t)^{\dagger},B_j\}\rho^{\rm eq}_{\rm E}]$ where $B_i$ is the operator of environments when $H_{\rm SE}$ can be expressed by $H_{\rm SE}=\sum_i S_iB_i$ where $S_i$ is the operator of a system.

We apply the Markovian approximation $\tau_{\rm C}\ll \tau_{\rm R}, \tau_{\rm C}\ll \Omega^{-1}$ to Eq.\ (\ref{qmmas}). By using $\tau = \tau_{\rm C}s$, $t=\tau_{\rm R}u, s \sim O(1), u \sim O(1)$, the integration in Eq.\ (\ref{qmmas}) becomes
 \begin{eqnarray}
& &\int_0^t d\tau  \tr_{\rm E} [H_{\rm SE}, [ e^{iH_0(\tau-t)/\hbar} H_{\rm SE}e^{-iH_0(\tau-t)/\hbar}, \rho^{\rm eq}_{\rm E}({\bm \pi}_{\tau})\rho(\chi,t) ]_{\chi}]_{\chi}\nonumber \\
&=& \int_0^{u\tau_{\rm R}/\tau_{\rm C}} \tau_{\rm C}ds  \tr_{\rm E} [H_{\rm SE}, [ e^{iH_0(-\tau_{\rm C}s)/\hbar} H_{\rm SE}e^{-iH_0(-\tau_{\rm C}s)/\hbar}, \rho^{\rm eq}_{\rm E}({\bm \pi}_{\tau_{\rm R}(u-s\tau_{\rm C}/\tau_{\rm R})})\rho(\chi,\tau_{\rm R}u) ]_{\chi}]_{\chi}\nonumber \\
&\simeq& \int_0^{\infty} d\tau  \tr_{\rm E} [H_{\rm SE}, [ e^{iH_0(-\tau)/\hbar} H_{\rm SE}e^{-iH_0(-\tau)/\hbar}, \rho^{\rm eq}_{\rm E}({\bm \pi}_{t})\rho(\chi,t) ]_{\chi}]_{\chi},
\end{eqnarray}
and $m(t)$ is negligible because, by using $dP(\tau)/d\tau=\Omega dP(v)/dv|_{v=\Omega \tau}$, 
 \begin{eqnarray}
m(t) &=& \tau_{\rm C}\Omega g^2\tr_B\left\{ \int_0^td\tau \mathcal L_{\rm SE}^{\chi} e^{-i\mathcal L_0(t-\tau)} \int_{\tau/\tau_{\rm C}}^{t/\tau_{\rm C}} ds \mathcal L_{\rm SE}^{\chi}(s) \left. \df{\mathcal P(v)}{v}\right|_{v=\Omega \tau} e^{i\mathcal L_0(t-\tau) }\rho_{\rm tot}(\chi,t) \right\} \nonumber \\
\end{eqnarray}
becomes much smaller than unity. Therefore, if we set the initial condition to $\rho_{\rm tot}(\chi,0) =\rho(\chi,0)\rho^{\rm eq}_{\rm E}({\bm \pi}_0)$, we obtain the Markovian master equation
\begin{eqnarray}
\df{}{t}\rho(\chi,t) &=& -\frac{i}{\hbar} [H_{\rm S}, \rho(\chi,t)] \nonumber \\
& &-\frac{g^2}{\hbar^2}\int_0^{\infty} d\tau  \tr_{\rm E} [H_{\rm SE}, [ e^{iH_0(-\tau)/\hbar} H_{\rm SE}e^{-iH_0(-\tau)/\hbar}, \rho^{\rm eq}_{\rm E}({\bm \pi}_{t})\rho(\chi,t) ]_{\chi}]_{\chi} \nonumber \\
\end{eqnarray}

\section{Adiabatic Markovian pumping: General expressions}
In this section, we briefly review the adiabatic Markovian pumping process under the condition $\Omega/\Gamma \ll 1$.
The argument in this section is parallel to that in Ref.\cite{Sagawa}.
Under this approximation, we can express the density matrix by the zero eigenvector that characterizes the steady state
as $\kket{\rho(\chi=0,t)} \simeq \kket{\lambda_+^0(\bm{\beta}(t))}$ where the subscript + represents the zero eigenvector and the superscript 0 represents the state without the counting field, i.e. $\chi=0$. 
Thus, the density matrix with the counting field $\chi$ can also be approximated by 
\begin{equation}
\kket{\rho(\chi,t)} \simeq c_+^{\chi}(t)e^{\Lambda^{\chi}_+(t)}\kket{\lambda_+^{\chi}(\bm{\beta}(t))} ,
\label{adi}
\end{equation}
where we have introduced a proportional constant that satisfies
\begin{equation}\label{adi_eq_c}
\dot{c}_+^{\chi}(t)=- c_+^{\chi}(t) \bbkk{l_+^{\chi}(\bm{\beta}(t))}{\dot{\lambda}_+^{\chi}(\bm{\beta}(t))},
\end{equation}
where we have used $\bbkk{l_+^{\chi}}{\lambda_+^{\chi}}=1$. 
Note that $\bbra{l_+^{\chi}}$ is reduced to $\bbra{1}$ for $\chi=0$, which means trace.

Equation (\ref{adi_eq_c}) is readily solvable as
\begin{eqnarray}\label{c_+}
c_+^{\chi}(t) &=& c_+^{\chi}(0) \exp{\left(- \int_0^td\tau\bbkk{l_+^{\chi}(\bm{\beta}(\tau))}{\dot{\lambda}_+^{\chi}(\bm{\beta}(\tau))} \right)} \nonumber \\
&=&  c_+^{\chi}(0) \exp{\left(- \int_{\mathcal{C}}\bbra{l_+^{\chi}(\bm{\beta})}d\kket{\lambda_+^{\chi}(\bm{\beta})} \right)} .
\end{eqnarray}
In the second line we have introduced the total differentiation $d$.
Substituting Eq.(\ref{c_+}) into Eq.(\ref{adi}) we obtain the cumulant-generating function $S(\chi,t)\equiv \ln{\bbkk{1}{\rho(\chi,t)}}$(see (\ref{rhotog})):
\begin{equation}\label{generating_fn2}
S(\chi,t) =- \int_{\mathcal{C}}\bbra{l_+^{\chi}(\bm{\beta})}d\kket{\lambda_+^{\chi}(\bm{\beta})} + \Lambda^{\chi}_+(t) + \ln{\frac{\bbkk{1}{\lambda_+^{\chi}(\bm{\beta}(t))}}{\bbkk{1}{\lambda_+^{\chi}(\bm{\beta}(0))}}} ,
\end{equation}
where the first, second, and the last terms on the RHS, respectively, correspond to the geometrical phase, the dynamical phase and the surface term.

Let us consider the energy transfer $\Delta q_t$ from the right reservoir to the system during time $t$. 
The average of $\Delta q_t$ can be calculated from the cumulant-generating function as $\langle \Delta q_t \rangle
=\partial S(\chi,t)/\partial (i\chi)|_{\chi=0}$.
Therefore, we obtain
\begin{equation}\label{adiabatic_current}
\expec{\Delta q_t} = \expec{\Delta q_t}_{\rm a}^{\rm g}+\expec{\Delta q_t}^{\rm d},
\end{equation}
where $\expec{\Delta q}_{\rm a}^{\rm g}$ represents the adiabatic pumping current in terms of the geometrical phase:
\begin{eqnarray}\label{adiabatic_geo_q}
\expec{\Delta q_t}_{\rm a}^{\rm g} &=& - \int_{\mathcal{C}}\bbra{l_+'(\bm{\beta})}d\kket{\lambda_+^0(\bm{\beta})} \nonumber \\
 &=&  -\int_0^td\tau \bbkk{l_+'(\tau)}{\dot{\lambda}_+^0(\tau)},
\end{eqnarray}
and $\expec{\Delta q_t}^{\rm d}$ is the adiabatic pumping current in terms of the dynamical phase:
\begin{eqnarray}\label{dyn_current}
\expec{\Delta q_t}^{\rm d} =\left. \pdf{\Lambda^{\chi}_+(t)}{(i\chi)}\right|_{\chi=0}= \int_{0}^td\tau \lambda_+'(\bm{\beta}(\tau)) .
\end{eqnarray}
where ${}'$ denotes the differentiation with respect to $\chi$.

Therefore, the adiabatic pumping current during the period $\tau_p$ can be written as
\begin{equation}\label{total_current}
J_{\rm a} = \frac{\expec{\Delta q_{\tau_p}}}{\tau_p} =J^{\rm g}_{\rm a}+J^{\rm d},
\end{equation}
where we have introduced 
\begin{eqnarray}
J^{\rm d} &=& \frac{1}{\tau_p}\int_0^{\tau_p}dt\lambda_+'(\bm{\beta}(t)), \label{jd}\\
J^{\rm g}_{\rm a} &=& -\frac{1}{\tau_p}\oint_{\mathcal{C}} \bbra{l_+'(\bm{\beta})}d\kket{\lambda_+^0(\bm{\beta})} \nonumber \\
&=& -\frac{1}{\tau_p}\iint_{\mathcal{S}} d\bbra{l_+'(\bm{\beta})}\wedge d\kket{\lambda_+^0(\bm{\beta})},
 \label{jg}
\end{eqnarray}
where $\iint_{\mathcal{S}}$ is the surface integral with the perimeter $\mathcal{C}$ and the integrand is called Berry curvature.
As shown in Appendix D, our adiabatic approximation is equivalent to that in Ref.\cite{Ren} if we apply this formulation to the spin-boson system.

\section{Adiabatic pumping for the spin-boson model}

In this appendix, we apply the general framework in the previous section to the spin-boson system (\ref{H_s&H_E}) and (\ref{H_{SE}})
to verify whether we can reproduce the results in Ref.\cite{Ren}.
In this case Eq.\ (\ref{Markov-K}) in Eq.\ (\ref{QME3}) is given by Eqs.\  (\ref{a1})--(\ref{a4}).
Furthermore, we also introduce
\begin{eqnarray}
b_2(\bm{\beta}(t)) &\equiv& - \left. \pdf{a_2^{\chi}({\bm \beta}(t))}{(i\chi)}\right|_{\chi=0}= -\hbar \omega_0\Gamma_{R}(1+ n_R(t)), \label{br2} \\ 
b_3(\bm{\beta}(t)) &\equiv & - \left. \pdf{a_3^{\chi}({\bm \beta}(t))}{(i\chi)}\right|_{\chi=0}= \hbar \omega_0 \Gamma_{R}n_R(t).
\label{br3}
\end{eqnarray}
To avoid complicated notations, we replace the parameter dependence through $\beta(t)$ by $t$.

From the differentiations of (\ref{lambda}) and (\ref{lefte}) we obtain
\begin{eqnarray}
\lambda'_{+} (t) &=& -\frac{a_1(t)b_2(t)+b_3(t)a_4(t)}{\lambda(t)}, \label{lambda_prime}\\
\bbra{l'_{+}(t)} &=& \begin{pmatrix}
0, & \frac{b_2(t)-b_3(t)}{\lambda(t)}
\end{pmatrix}. \label{leftprime}
\end{eqnarray}
Substituting Eqs.\ (\ref{a1}),(\ref{a4}),(\ref{br2}),(\ref{br3}) into (\ref{lambda_prime}) we can rewrite  
\begin{eqnarray}
\lambda'_{+} (t) = \frac{\hbar \omega_0 \Gamma_L\Gamma_R(n_L(t)-n_R(t))}{\lambda(t)} .
\end{eqnarray}
Substituting this into Eq.(\ref{jd}) we obtain the dynamical current
\begin{eqnarray}\label{Rendyn}
J^{\rm d} = \frac{\hbar \omega_0}{\tau_p}\int_0^{\tau_p}dt \frac{\Gamma_L\Gamma_R(n_L(t)-n_R(t))}{\lambda(t)} ,
\end{eqnarray}
which is equivalent to Eq.(13) of Ref.\cite{Ren}. In the case of a symmetric junction under the environments $\Gamma = \Gamma_L=\Gamma_R$ without average bias, Eq.(\ref{Rendyn}) can be rewritten as
\begin{eqnarray}\label{Rendynsym}
J^{\rm d} = \frac{\hbar \omega_0\Gamma }{2\tau_p}\int_0^{\tau_p}dt \frac{n_L(t)-n_R(t)}{1+n_L(t)+n_R(t)} = 0 ,
\end{eqnarray}
To derive the final equality of Eq.\ (\ref{Rendynsym}) we use the idea that $n_L$ and $n_R$ are sinusoidal functions of time, and thus, $n_L/(1+n_L+n_R)$ sweeps an identical area to $n_R/(1+n_L+n_R)$ during a period. 

On the other hand, let us rewrite the integrand in Eq.(\ref{jg}) as  
\begin{equation}\label{integrand}
-d\bbra{l_+'(\bm{\beta})}\wedge d\kket{\lambda_+^0(\bm{\beta})} = dT_LdT_R \left( \pdf{}{T_R}\bbra{l_+'(\bm{\beta})}\pdf{}{T_L}\kket{\lambda_+^0(\bm{\beta})} - 
\pdf{}{T_L}\bbra{l_+'(\bm{\beta})}\pdf{}{T_R}\kket{\lambda_+^0(\bm{\beta})}\right)
\end{equation}
where we have used $dT_L \wedge dT_R = + dT_LdT_R$.
Because of (\ref{leftprime}) the only relevant term is the second component in the above equation.
From the straightforward calculation, we can rewrite Eq.(\ref{integrand}) as
\begin{eqnarray}
& &-d\bbra{l_+'(\bm{\beta})}\wedge d\kket{\lambda_+^0(\bm{\beta})}\\
&=& dT_LdT_R\frac{2\hbar \omega_0(k_{\rm B}\beta_L^2)(k_{\rm B}\beta_R^2)\Gamma_L\Gamma_R(\Gamma_L+\Gamma_R)}{\lambda^3}
\pdf{n_L}{\beta_{L}}\pdf{n_R}{\beta_{R}}.
\end{eqnarray}
Introducing 
$C_{\nu}\equiv \pdf{n_{\nu}}{T_{\nu}}=k_{\rm B}\beta_{\nu}^2\hbar\omega_0e^{\beta_{\nu}\hbar\omega_0}n_{\nu}^2$
Eq.(\ref{jg}) is thus reduced to
\begin{eqnarray}
J^{\rm g}_{\rm a} = \frac{\hbar \omega_0}{\tau_p}\iint_{\mathcal{S}}d\tau_Ld\tau_R\frac{2C_LC_R\Gamma_L\Gamma_R(\Gamma_L+\Gamma_R)}{\lambda^3} .
\end{eqnarray}
Thus, we reproduce Eqs.\  (14) and (15) of Ref.\cite{Ren}.

\section{Asymptotic expansion}
In this section, we prove the asymptotic expansion of the density matrix appearing in Eqs.\ (\ref{rhoasy})--(\ref{rhoasyco}) and the non-adiabatic pumping current in Eqs.\ (\ref{pzen})--(\ref{pb2}) in the limit $\Omega/\Gamma \rightarrow 0$.

Suppose that $\rho_{11}(t)$ is given by Eq.(\ref{rhot}). Introducing $\tilde{u}(t) \equiv \int_0^tdt' (n_L(\Omega t') + n_R(\Omega t'))$, $\rho_{11}(t)$ can be represented by
\begin{eqnarray}
\rho_{11}(t) = \left(\rho_{11}(0) - \frac{1}{2} \right) e^{-2\Gamma (t+\tilde{u}(t))} +\frac{1}{2}-\Gamma  e^{-2\Gamma (t+\tilde{u}(t))}\int_0^{t}dt' e^{2\Gamma (t'+\tilde{u}(t'))}.
\end{eqnarray}
The pumping current (\ref{exccurrent}) can be written as
\begin{eqnarray}
J^{\rm E}_{\rm na}  = -\frac{1}{\tau_p}\int_0^{\tau_p}dt \frac{b_2(t)-b_3(t)}{\lambda(t)}(\lambda(t)\rho_{11}(t) -a_1(t) ).
\end{eqnarray}
Let us introduce the dimensionless variables $\theta = \Omega t$ and $u(\theta) \equiv \Omega \tilde{u}(t(\theta))=\int_0^{\theta}d\theta' (n_L(\theta') + n_R(\theta'))$. Then we can write
\begin{eqnarray}
J^{\rm E}_{\rm na} &=& -\frac{1}{2\pi}\int_0^{2\pi}d\theta (b_2(\theta)-b_3(\theta))\left(\rho_{11}(\theta) -\frac{a_1(\theta)}{\lambda(\theta) }\right), \nonumber \\\label{pump} \\
\rho_{11}(\theta)&=&\left(\rho_{11}(0) - \frac{1}{2} \right) e^{-s\theta -s u(\theta)} +\frac{1}{2} -\frac{s}{2}  e^{-s\theta -su(\theta)}\int_0^{\theta}d\theta' e^{s\theta' +su(\theta') }, \label{rho}
\end{eqnarray}
where we introduce $s \equiv 2\Gamma/\Omega$. 
Let us consider the asymptotic behavior of $\rho_{11}(\theta)$ in the limit $s \rightarrow \infty$, i.e. $\Omega/\Gamma \rightarrow 0$. The last term on the RHS of (\ref{rho}) with $s(\theta-\theta') = \xi$ can be rewritten as
\begin{eqnarray}
& &s  e^{-s\theta -s u(\theta)}\int_0^{\theta}d\theta' e^{s\theta' +s u(\theta')}\nonumber \\
&= &e^{-s u(\theta)}  \int_0^{s\theta}d\xi e^{-\xi +s u(\theta-\xi/s)} \nonumber \\
&=& e^{-s u(\theta)} \int_0^{s\theta}d\xi e^{ -\xi } \sum_{n=0}^{\infty}\frac{s^n}{n!} \sum_{k_1,..., k_n =0}^{\infty}\frac{u^{(k_1)}(\theta)\cdots u^{(k_n)}(\theta)}{k_1!\cdots k_n!}\left( \frac{-\xi}{s} \right)^{k_1+\cdots +k_n}. \label{expan}
\end{eqnarray}
From an identity of the gamma function
\begin{eqnarray}
\int_0^{s\theta}d\xi e^{ -\xi } \xi^{k_1+\cdots +k_n} = \Gamma(k_1+\cdots k_n+1)-\int_{s\theta}^{\infty}d\xi e^{ -\xi } \xi^{k_1+\cdots +k_n}, \label{gamma}
\end{eqnarray}
the asymptotic expansion of the incomplete gamma function
\begin{eqnarray}
\int_{x}^{\infty}dte^{-t}t^{a-1}
&\simeq& e^{-x}x^{a-1}\left(1+\frac{a-1}{x}+\frac{(a-1)(a-2)}{x^2}+ \cdots \right. \nonumber \\
& & \left. \hspace{2cm} +\frac{(a-1)(a-2)\cdots(a-m)}{x^m}+\cdots\right), \nonumber 
\end{eqnarray}
and the Leibniz rule
\begin{eqnarray}
\sum_{k_1,\cdots k_n =0}^{\infty}\frac{(k_1+\cdots+k_n)!}{k_1!\cdots k_n!} g_1^{(k_1)}(x)\cdots g_n^{(k_n)}(x)= \sum_{k=0}^{\infty}\frac{\partial^k}{\partial x^k}(g_1(x)\cdots g_n(x)),
\end{eqnarray}
Eq.\ (\ref{expan}) becomes
\begin{eqnarray}
1+ \sum_{k=1}^{\infty}\left. \frac{\partial^k}{\partial x^{k}} \exp{\left( s\sum_{j=1}^{\infty} \frac{u^{(j)}(\theta)}{j!}\left(\frac{-x}{s}\right)^j\right)}\right|_{x=0}+ e^{-s\theta-su(\theta)}\sum_{m=0}^{\infty} \left. \frac{\partial^m}{\partial y^m}e^{su(-y/s)}\right|_{y=0}.\nonumber 
\end{eqnarray}
Thus, we obtain the expression for $\rho_{11}(\theta)$ as
\begin{eqnarray}\label{rhoall}
\rho_{11}(\theta)&=&\left(\rho_{11}(0) - \frac{1}{2} \right) e^{-s\theta -s u(\theta)}-\frac{1}{2}\sum_{k=1}^{\infty}\left. \frac{\partial^k}{\partial x^{k}} \exp{\left( s\sum_{j=1}^{\infty} \frac{u^{(j)}(\theta)}{j!}\left(\frac{-x}{s}\right)^j\right)}\right|_{x=0}\nonumber \\
& & +\frac{e^{-s\theta-su(\theta)}}{2}\sum_{m=0}^{\infty} \left. \frac{\partial^m}{\partial y^m}e^{su(-y/s)}\right|_{y=0}.
\end{eqnarray}
If Eq.\ (\ref{rhoall}) is expanded up to the second order of $1/s$ and $s \theta \gg 1$, we obtain (\ref{rhoasy})-(\ref{rhoasyco}). 

Substituting Eq.\ (\ref{rhoall}) into (\ref{pump}) we obtain the pumping current
\begin{eqnarray}\label{Jall}
	J &=& \frac{1}{2\pi}\int_0^{2\pi}d\theta (b_2(\theta)-b_3(\theta)) \left(\frac{1}{2}\sum_{k=1}^{\infty}\left. \frac{\partial^k}{\partial x^{k}} \exp{\left( s\sum_{j=1}^{\infty} \frac{u^{(j)}(\theta)}{j!}\left(\frac{-x}{s}\right)^j\right)}\right|_{x=0}+\frac{a_1(\theta)}{\lambda(\theta) }\right) \nonumber \\
& & - \frac{1}{2\pi}\int_0^{2\pi}d\theta (b_2(\theta)-b_3(\theta))e^{-s\theta -s u(\theta)} \left[ \rho_{11}(0) - \frac{1}{2} + \frac{1}{2}\sum_{m=0}^{\infty}\left. \frac{\partial^m}{\partial y^m}e^{su(-y/s)}\right|_{y=0}\right]. \label{kekka}
\end{eqnarray}
Let us denote $f(\theta)$ for the second term on the RHS of Eq.\ (\ref{Jall}) except for the exponential factor, which satisfies $f(\theta+2\pi)=f(\theta)$. Let $\delta u(\theta)$ be the fluctuation part of $u(\theta)$: 
\begin{eqnarray}
\delta u(\theta) &\equiv& u(\theta) -\bar{u}\theta, \\
\bar{u} &\equiv& \frac{u(2\pi)}{2\pi}.
\end{eqnarray}
Then we can write
\begin{eqnarray}
\int_0^{2\pi}d\theta f(\theta)e^{-s\theta -s u(\theta)}
&=&\left( \int_0^{\infty} -\int_{2\pi}^{\infty}\right)d\theta f(\theta)e^{-s(1+\bar{u})\theta -s \delta u(\theta)} \nonumber \\
&=&(1-e^{-2\pi s(1+\bar{u})})\int_0^{\infty} d\theta f(\theta)e^{-s\theta -s u(\theta)}.
\end{eqnarray}
This integration can be rewritten as
\begin{eqnarray}\label{fseki}
& &\int_0^{\infty} d\theta f(\theta)e^{-s\theta -s u(\theta)} \nonumber \\
&=& \int_0^{\infty} d\theta \left(\sum_{m=0}^{\infty}\frac{f^{(m)}(0)}{m!}\theta^m\right) \left(\sum_{n=0}^{\infty}\frac{(-su(\theta))^n}{n!}\right)e^{-s\theta} \nonumber \\
&=& \sum_{m,n=0}^{\infty}\frac{f^{(m)}(0)}{m!}\frac{(-s)^n}{n!}\sum_{k_1,..., k_n =0}^{\infty}\frac{u^{(k_1)}(0)\cdots u^{(k_n)}(0)}{k_1!\cdots k_n!}\frac{\Gamma(m+k_1+\cdots+k_n+1)}{s^{m+k_1+\cdots+k_n+1}} \nonumber \\
&=& \frac{1}{s}\sum_{k=0}^{\infty}\left. \frac{\partial^k}{\partial z^k}(f(z/s)e^{-su(z/s)})\right|_{z=0} 
\end{eqnarray}
Substituting Eq.\ (\ref{fseki}) into Eq.\ (\ref{Jall}) we obtain
\begin{eqnarray}
	J &=& \frac{1}{2\pi}\int_0^{2\pi}d\theta (b_2(\theta)-b_3(\theta))\left(\frac{1}{2}\sum_{k=1}^{\infty}\left. \frac{\partial^k}{\partial x^{k}} \exp{\left( s\sum_{j=1}^{\infty} \frac{u^{(j)}(\theta)}{j!}\left(\frac{-x}{s}\right)^j\right)}\right|_{x=0}+\frac{a_1(\theta)}{\lambda(\theta) }\right) \nonumber \\
& & -\frac{1-e^{-2\pi s(1+\bar{u})}}{2\pi s}\sum_{k=0}^{\infty}\left. \frac{\partial^k}{\partial z^k}(f(z/s)e^{-su(z/s)})\right|_{z=0} , \label{jzen} \\
f(\theta) &=& (b_2(\theta)-b_3(\theta))\left[ \rho_{11}(0) - \frac{1}{2} +\frac{1}{2}\sum_{m=0}^{\infty} \left. \frac{\partial^m}{\partial y^m}e^{su(-y/s)}\right|_{y=0} \right]. \label{nakami}
\end{eqnarray}
If this formula is expanded up to the second order of $1/s$, we reach Eq.\ (\ref{pzen})-(\ref{pb2}).

\section{Derivation of fluctuation theorems}
In this appendix, we explain the detailed derivation of two types of extended heat fluctuation theorems for slowly modulated temperatures. In Appendix F.1, we discuss the extended heat fluctuation theorem in the limit $\tau_p\omega_0\to \infty$. In Appendix F.2, we discuss the extended fluctuation theorem for a large amount of transferred energy.

\subsection{Derivation for $\tau_p\omega_0 \rightarrow \infty$}
In this subsection, we derive the fluctuation theorem Eq.\ (\ref{tft}) under the condition that the period of the modulation $\tau_p \omega_0$ is sufficiently large.

The probability distribution of the transferred energy $P(\Delta q_{\tau_p})$ during a period $\tau_p$ is expressed by the Fourier transform $P(\Delta q_{\tau_p}) = \int d\chi e^{-i\chi \Delta q_{\tau_p} + S(\chi,\tau_p)}/2\pi$. By introducing the current variable $\xi(t) = \Delta q_t$ and the differential cumulant-generating function $s_t(\chi)$:
\begin{eqnarray}
s_t(\chi) &=& \lambda_+^{\chi}(t) + \tilde v^{\chi}(t),
\end{eqnarray}
with
\begin{eqnarray}
\tilde v^{\chi}(t) &=& - \bbra{l_+^{\chi}(\bm \beta(t))} \df{}{t} \kket{\lambda_+^{\chi}(\bm \beta(t))},
\end{eqnarray}
the probability distribution is represented as
\begin{eqnarray}
P(\xi) &=& \int_{-\infty}^{\infty}\frac{d\chi}{2\pi}e^{-\tau_p\omega_0 F(\chi,\xi)}, \label{pdx}
\end{eqnarray}
where
\begin{eqnarray}
F(\chi,\xi) &=& \frac{1}{\tau_p}\int_0^{\tau_p}dt \{ i\chi \xi(t) - s_t(\chi) \}/\omega_0.
\end{eqnarray}
Let us introduce 
\begin{eqnarray}
I(\xi) = \max_{\chi}\frac{1}{\tau_p}\int_0^{\tau_p}dt \{ i\chi \xi(t) - \lambda_+^{\chi} (t) \}/\omega_0, \label{ratex}
\end{eqnarray}
which is reduced to the usual rate function when temperatures do not depend on the time. Let $\chi^*=\chi^*(\overline{\xi})$ be what maximizes Eq.\ (\ref{ratex}), i.e., which satisfies $i\overline{\xi} = \partial \overline{\lambda_+^{\chi}(t)}/\partial \chi |_{\chi^*}$. It should be noted that under the non-stationary modulation, the GC symmetry $\lambda_+^{\chi}(t) = \lambda_+^{-\chi+i\alpha(t)}(t)$ gives the relation
\begin{eqnarray}
I(\xi)-I(-\xi) = -\overline{\alpha(t) \xi(t)}. \label{rate}
\end{eqnarray}
By extracting $I(\xi)$ from Eq.\ (\ref{pdx}), $P(\xi)$ is rewritten as 
\begin{eqnarray}
P(\xi) &=& e^{-\tau_p\omega_0 I(\xi)}\int_{-\infty}^{\infty}\frac{du}{2\pi}y(u,\xi)e^{-\tau_p\omega_0 u^2/2}, \label{pdu}
\end{eqnarray}
where we have introduced $u = \chi \sqrt{2(F(\chi,\xi)-I(\xi))}/|\chi|$ and $y(u,\xi) = d\chi/du = u(\partial F(\chi,\xi)/\partial \chi)^{-1}$.

In the limit $\tau_p \omega_0 \rightarrow \infty$, Eq.\ (\ref{pdu}) can be evaluated near $u=0$. From the expansion $y(u,\xi) = \sum_{n=0}z_{n}(\xi) u^n/n!$ with $z_n(\xi) = \partial^n y(u,\xi)/\partial u^n |_{u=0}$, Eq.\ (\ref{pdu}) can be rewritten as 
\begin{eqnarray}
P(\xi) &\simeq& e^{-\tau_p\omega_0 I(\xi)}\frac{1}{\sqrt{2\pi \tau_p\omega_0}}\left[z_0(\xi) + z_2(\xi)\frac{1}{2\tau_p\omega_0} + O\left( (\tau_p\omega_0)^{-2} \right) \right]. \label{tas}
\end{eqnarray}
Here, let $y(u,\xi)$ be expanded in $\tau_p^{-1}$. For this purpose, introducing a dimensionless quantity $ v^{\chi}(\theta) \equiv \tilde v^{\chi}(\theta/\Omega)\tau_p$ with $\theta = \Omega t$, we rewrite  $g_{\xi}(\chi) \equiv F(\chi,\xi)-I(\xi) $ as
\begin{eqnarray}
g_{\xi}(\chi) = A_{\xi}(\chi) -\frac{1}{\tau_p\omega_0} \overline{v^{\chi}(\theta) }, \label{gex}
\end{eqnarray}
where we have used 
\begin{eqnarray}
A_{\xi}(\chi)=\frac{1}{\tau_p}\int_0^{\tau_p}dt \{ i(\chi-\chi^*) \xi(t) -(\lambda_+^{\chi}(t)-\lambda_+^{\chi^*} (t) )\}/\omega_0,
\end{eqnarray}
which obviously satisfies $A_{\xi}(\chi^*) = 0$ and $A_{\xi}'(\chi^*)=0$, where $'$ denotes a $\chi$-derivative.
Let us introduce $\chi_0$ satisfying $g_{\xi}(\chi_0) =0$ corresponding to $u=0$ in Eq.\ (\ref{pdu}). Then, from Eq.\ (\ref{gex}), $\chi_0$ can be obtained as the series of $(\tau_p\omega_0)^{-1/2}$;
\begin{eqnarray}
\chi_0 \simeq \chi^* + \frac{b_1}{\sqrt{\tau_p \omega_0}} +\frac{b_2}{\tau_p\omega_0}+ O\left( (\tau_p\omega_0)^{-3/2}\right) \label{chiex}
\end{eqnarray}
with $b_1= \sqrt{\frac{2\overline{v^{\chi^*}(\theta) }}{A_{\xi}''(\chi^*)}}$ and $b_2 =\frac{\partial \overline{ v^{\chi}(\theta) }/\partial \chi|_{\chi^*}}{A_{\xi}''(\chi^*)}-\frac{A_{\xi}'''(\chi^*)\overline{ v^{\chi^*}(\theta) }}{3A_{\xi}''(\chi^*)^2} $.

It should be noted that $z_0(\xi)$ in Eq.\ (\ref{tas}) becomes zero because the denominator of $z_0(x)$ is $g'(\chi_0) = A''(\chi^*)b_1(\tau_p\omega_0)^{-1/2}+\cdots \neq 0$ and the numerator $g(\chi_0)=0$. 
Thus, the dominant contribution $z_2(\xi)$ can be expanded as
\begin{eqnarray}
z_2(\xi) = - \frac{g_{\xi}''(\chi_0)}{g_{\xi}'(\chi_0)^3} \simeq \frac{(\tau_p\omega_0)^{3/2}}{\{(2\overline{ v^{\chi^*}(\theta) })^3A_{\xi}''(\chi^*)\}^{1/2}}\left[ 1+ O\left( (\tau_p\omega_0)^{-1} \right) \right] \label{z2ex}
\end{eqnarray}
with the aid of Eq.\ (\ref{chiex}). By substituting Eq.\ (\ref{z2ex}) into Eq.\ (\ref{tas}), $P(\xi)$ is rewritten as
\begin{eqnarray}
P(\xi) &\simeq& e^{-\tau_p\omega_0 I(\xi)}\frac{1}{2\sqrt{2\pi }}\left[\frac{1}{\{(2\overline{ v^{\chi^*}(\theta) })^3A_{\xi}''(\chi^*)\}^{1/2}} + O\left( (\tau_p\omega_0)^{-1} \right) \right]. \label{pexchi}
\end{eqnarray}
Hence, Eqs.\ (\ref{rate}) and (\ref{pexchi}) give Eq.\ (\ref{tft}) used in the main text. 

\subsection{Derivation for $N \rightarrow \infty$}
In this subsection, we derive the fluctuation theorem Eq.\ (\ref{sft}) by using coupled master equations. Here, we assume that the number of the transferred charge $N$ is sufficiently large.

Let $\hat \rho^{q}(t)$ be the Fourier transform of  $\rho(\chi,t)$ defined by
\begin{eqnarray}
\hat \rho^{q}(t) = \int_{-\infty}^{\infty}\frac{d\chi}{2\pi}e^{-i\chi q}\rho(\chi,t),
\end{eqnarray}
where $q$ is the transferred energy at $t$. 
Because $\rho(\chi,t)^{\dagger} = \rho(-\chi,t)$, $\hat \rho^q(t)$ is a Hermitian matrix. Therefore, $\hat \rho^q(t)$ can be used for spectral decomposition:
\begin{eqnarray}
\hat \rho^q(t) = \sum_{m_t}r_{m_t}^q(t) \ket{m_t}\bra{m_t}.
\end{eqnarray}
With the aid of $r_m^q$, the quantum master equation is given by
\begin{eqnarray}\label{2dmas}
\dot r^q_m(t) = \sum_{m'}\int dq' W_t(m,m'|q,q') r_{m'}^{q'}(t),
\end{eqnarray}
where $W_t(m,m'|q,q') \equiv \tr\{ \ket{m}\bra{m}\int d\chi/2\pi e^{-i\chi(q-q')}\mathcal K^{\chi}(\bm{\beta}(t)) \ket{m'}\bra{m'}\}$ is the transition rate from a state $(m',q')$ to a state $(m,q)$ at $t$. 

In a spin-boson system, the master equation (\ref{2dmas}) is reduced to a set of coupled equations for $r_0^q(t)$ and $r_1^q(t)$ as
\begin{eqnarray}\label{2dmas}
\dot r^q_0(t) &=& -\Gamma \{ n_L(t) + n_R(t) \}r_0^q(t) +\Gamma n_L(t)r_1^q(t) +\Gamma n_R(t)r_1^{q-\hbar \omega_0}, \\
\dot r^q_1(t) &=& \Gamma \{1+n_L(t)\}r_0^q(t) +\Gamma \{1+n_R(t)\}r_0^{q+\hbar \omega_0}-\Gamma \{ 2+n_L(t) + n_R(t) \}r_1^q(t), 
\end{eqnarray}
where $r_0^q = \int d\chi/2\pi e^{-i\chi q}\rho_{00}(\chi,t), r_1^q = \int d\chi/2\pi e^{-i\chi q}\rho_{11}(\chi,t)$ because the diagonal components of the density matrix are independent of the non-diagonal components.
If the total transferred energy is $q = N \hbar \omega_0$, the forward and backward paths are given by
\begin{eqnarray}
r_1^{0} \overset{\Gamma n_R}{\underset{\Gamma (1+n_R)}{\rightleftarrows}} r_0^{+1} \overset{\Gamma(1+ n_L)}{\underset{\Gamma n_L}{\rightleftarrows}} r_1^{+1} \rightleftarrows \cdots \rightleftarrows r_1^{+N},
\end{eqnarray}
where each transition rate is written on the arrow.

In the case of no modulation of parameters, the ratio of the probability of the forward path $P_F(+N\hbar \omega_0)$ to the probability of the backward path $P_B(-N\hbar \omega_0)$ gives the conventional fluctuation theorem
\begin{eqnarray}
\ln \frac{P_F(+N\hbar \omega_0)}{P_B(-N\hbar \omega_0)} = \ln \left( \frac{\Gamma n_R(1+n_L)}{\Gamma (1+n_R)n_L }\right)^N = N \hbar \omega_0(\beta_L-\beta_R).
\end{eqnarray}

In the case of finite modulation of parameters, we use the method of Ref.\cite{Esposito2}. 
First, we divide aa interval $[0,\tau_p]$ into short intervals $[\tau_{j-1},\tau_j]$, where $\tau_j = j\Delta \tau = j 2\tau_p/(2N+1)$. The transferred energy at $\tau=\tau_j$ is $q_j = j\hbar \omega_0$. Each state $r_m^q$ stays between $\Delta \tau/2$. 
Then, the probability for the forward trajectory $\mu_F[q]$ is given by
\begin{eqnarray}
\mu_F[q] &=& r_1(0) \left[\prod_{j=1}^N e^{ \int_{\tau_{j-1}}^{\tau_{j-1}+\Delta \tau/2} d\tau' W_{\tau'}(1,1|q_{j-1},q_{j-1})}W_{\tau_{j-1}+\Delta \tau/2}(0, 1|q_j,q_{j-1}) \right. \nonumber \\
& & \left. \times e^{ \int_{\tau_{j-1}+\Delta \tau/2}^{\tau_j} d\tau' W_{\tau'}(0,0|q_j,q_j)}W_{\tau_{j}}(1, 0|q_j,q_j)\right]e^{ \int_{\tau_{N}}^{\tau_p} d\tau' W_{\tau'}(1,1|q_N,q_N)},
\end{eqnarray}
where each exponential is the probability of staying in the state. On the other hand, by using the time-reversal transformation $\tau \rightarrow \tilde \tau = \tau_p -\tau, q_j \rightarrow \tilde q_j = q_{N-j}$, the probability of the backward trajectory $\mu_B[\tilde q]$ is given by
\begin{eqnarray}
\mu_B[\tilde q] &=& r_1(\tau_p) \left[\prod_{k=1}^N e^{ \int_{\tau_{k-1}}^{\tau_{k-1}+\Delta \tau/2} d\tau' W_{\tau'}(1,1|q_{k-1},q_{k-1})}W_{\tau_{k-1}+\Delta \tau/2}(0, 1|q_{k-1},q_{k-1}) \right. \nonumber \\
& & \left. \times e^{ \int_{\tau_{k-1}+\Delta \tau/2}^{\tau_k} d\tau' W_{\tau'}(0,0|q_{k-1},q_{k-1})}W_{\tau_k}(1, 0|q_{k-1},q_k)\right]e^{ \int_{\tau_{N}}^{\tau_p} d\tau' W_{\tau'}(1,1|q_N,q_N)}.
\end{eqnarray}
Therefore, the ratio of the probabilities of the two trajectories is reduced to
 \begin{eqnarray}
\ln \frac{\mu_F[q]}{\mu_B[\tilde q] }&=& \ln \frac{r_1(0)}{ r_1(\tau_p)} + \sum_{j=1}^{N} \ln \frac{W_{\tau_{j-1}+\Delta \tau/2}(0, 1|q_j,q_{j-1})W_{\tau_{j}}(1, 0|q_j,q_j)}{ W_{\tau_{j-1}+\Delta \tau/2}(0, 1|q_{j-1},q_{j-1}) W_{\tau_j}(1, 0|q_{j-1},q_j)} \nonumber \\
&=& \hbar \omega_0 \sum_{j=1}^{N}\{\beta_L(\tau_j)-\beta_R(\tau_j) \} +  \sum_{j=1}^{N} \ln \frac{e^{\beta_L(\tau_{j-1}+\Delta \tau/2)\hbar \omega_0}-1}{e^{\beta_R(\tau_{j-1}+\Delta \tau/2)\hbar \omega_0}-1}\frac{e^{\beta_R(\tau_j)\hbar \omega_0}-1}{e^{\beta_L(\tau_j)\hbar \omega_0}-1}, \nonumber \\ \label{caltra}
\end{eqnarray}
where we assume that $r_1(0) = r_1(\tau_p)$. From the periodicity $\bm{\beta}(t+\tau_p)=\bm{\beta}(t)$, the parameters can be used for Fourier expansoin, e.g. $\beta_L(\tau_j) = \sum_mc_me^{im 4\pi j/(2N+1)}$. Then, we obtain
 \begin{eqnarray}
\sum_{j=1}^{N}\beta_L(\tau_j) \simeq N\bar \beta_L + \frac{\bar \beta_L-\beta_L(0)}{2} -\frac{5\pi}{4N}\left.\df{\beta_L(\theta)}{\theta}\right|_{\theta=0} + O\left(\frac{1}{N^2}\right).
\end{eqnarray}
With the aid of Eq.\ (\ref{caltra}) and the replacements $P_F(q)=\mu_F[q]$ and $P_B(-q)=\mu_B[\tilde q]$, we obtain Eq.\ (\ref{sft}).

\end{document}